\documentclass{CSML}

\def\dOi{13(1:8)2017}
\lmcsheading%
{\dOi}
{1--33}
{}
{}
{Jul.~15, 2010}
{Mar.~17, 2017}
{}

\newcommand\VL[1]{#1} 		
\newcommand\VML[1]{#1} 		
\newcommand\VM[1]{}		
\newcommand\VSM[1]{} 	
\newcommand\VS[1]{} 	

\newcommand\couic[1]{} 	

\usepackage{amsfonts}
\usepackage{amsmath}
\usepackage{latexsym}
\usepackage{amssymb}
\usepackage{stmaryrd}
\usepackage{hyperref}

\def\lra{\longrightarrow} 
\def\lla{\longleftarrow}

\newcommand{\ket}[1]{| #1 \rangle}

\makeatletter
\DeclareRobustCommand*\cal{\@fontswitch\relax\mathcal}
\makeatother

\begin{document}

\title[Lineal: A linear-algebraic $\lambda$-calculu]{Lineal: A linear-algebraic $\lambda$-calculus}

\author[P.\ Arrighi]{Pablo Arrighi\rsuper a}
\address{{\lsuper a}Universit\'e de Grenoble, Laboratoire LIG, UMR 5217, 220 rue
  de la Chimie, 38400 Saint-Martin d'H\`eres, France, and Universit\'e
  de Lyon, Laboratoire LIP, UMR 5668, 46 all\'ee d'Italie 69007 Lyon,
  France.}
\email{pablo.arrighi@imag.fr}

\author[G.~Dowek]{Gilles Dowek\rsuper b}
\address{{\lsuper b}INRIA, 23 avenue d'Italie, CS 81321, 75214 Paris Cedex 13, France.}
\email{gilles.dowek@inria.fr}


\begin{abstract}
We provide a computational definition of the notions of vector space
and bilinear functions. We use this result to introduce a minimal
language combining higher-order computation and linear algebra. This
language extends the $\lambda$-calculus with the possibility to make
arbitrary linear combinations of terms $\alpha.{\bf t}+\beta.{\bf
u}$. We describe how to ``execute'' this language in terms of a few
rewrite rules, and justify them through the two fundamental
requirements that the language be a language of linear operators, and
that it be higher-order. We mention the perspectives of this work in
the field of quantum computation, whose circuits we show can be easily
encoded in the calculus. Finally, we prove the confluence of the
entire calculus.
\end{abstract}

\maketitle

\section{Motivations}

Knuth and Bendix have described a method to transform an equational
theory into a rewrite system \cite{KnuthBendix}. In this paper, we
show that this can be achieved for the theory of vector spaces.  This
yields a computational definition of the notion of vector space. We
then use this definition to merge at a fundamental level higher-order
computation in its simplest and most general form, the untyped
$\lambda$-calculus, together with linear algebra. We see this
{\em Linear-algebraic $\lambda$-calculus} (also referred to as {\em Lineal}
for short) as a platform for various applications, such as
non-deterministic, probabilistic and quantum computation --- each of
these applications probably requiring their own type systems.

This journal paper integrates the contributions from three already published conference papers \cite{arrighidowek1,arrighidowek2,arrighidowek3}. There has been a number of recent works surrounding these papers, whose presentation we postpone till Section \ref{currentworks}. The emphasis of the present introduction is on the original motivations behind this calculus; in the same way that the emphasis of the present paper is on providing an integrated, coherent, comprehensive presentation of the calculus without further add-ons. 

\subsection{Quantum programming languages} 
\VML{Over
the last two decades, the discovery of several great algorithmic
results \cite{Deutsch,Shor,Grover} has raised important expectations
in the field of quantum computation. Somewhat surprisingly however
these results have been expressed in the primitive model of quantum
circuits -- a situation which is akin to that of classical computation
in the 1950s. Over the last few years a number of researchers have
sought to develop quantum programming languages as a
consequence. Without aiming to be exhaustive and in order to
understand where the perspectives of this work come in, it helps to
classify these proposals according to ``how classical'' versus ``how
quantum'' they are \cite{Valiron}.} There are two ways a quantum mechanical system may evolve: according
to a {unitary transformation} or under a {measurement}. The former is
often thought of as ``purely quantum'': it is deterministic and will
typically be used to obtain quantum superpositions of base
vectors. The latter is probabilistic in the classical sense, and will
typically be used to obtain some classical information about a quantum
mechanical system, whilst collapsing the system to a mere base
vector.

\VML{Note that these are only typical uses: it is well-known that
one can simulate any unitary transformation by series of generalized
measures on the one hand, and reduce all measures to a mere projection
upon the canonical basis at the end of a computation on the other
hand. It remains morally true nonetheless that {measurement-based}
models of quantum computation tend to hide quantum superpositions
behind a classical interface, whilst the {unitary-based} models of
quantum computation tend to consider quantum superpositions as
legitimate expressions of the language, and sometimes even seek to
generalize their effects to control flow.}

\VML{Therefore one}\VS{One} may say that measurement-based models of quantum computation -- whether reliant
upon teleportation \cite{Nielsen}, state transfer \cite{Perdrix} or
more astonishingly graph states \cite{Briegel} -- lie on one extreme, as they keep the ``quantumness''
to a minimum. \VML{ 

}A more balanced approach is to allow for both unitary 
transformations and quantum measurements. Such models can be said to
formalize the existing algorithm description methods to a strong
extent: they exhibit quantum registers upon which quantum circuits may
be applied, together with classical registers and programming
structures in order to store measurements results and control the
computation \cite{Selinger}. For this reason they are the more
practical route to quantum programming. Whilst this juxtaposition of
``quantum data, classical control'' has appeared ad-hoc and
heterogeneous at first, functional-style approaches together with
linear type systems \cite{Valiron,Altenkirch} have ended up producing
elegant quantum programming languages.

Finally we may evacuate measures altogether -- leaving them till the end of the computation
and outside the formalism.  This was the case for instance in
\cite{VanTonder1,VanTonder2}, but here the control structure remained
classical. 

In our view, such a language becomes even more interesting
once we have also overcome the need for any additional classical
registers and programming structures, and aim to draw the full
consequence of quantum mechanics: ``quantum data, quantum control''.
After all, classical control can be viewed as a particular case of quantum control. 
Moreover, avoiding this distinction leads to a simpler language, exempt of the separation between classical and quantum expressions. Finally recent results suggest that quantum control may turn out to be more efficient that classical control in the presence of Black-box algorithms \cite{OreshkovCostaBrukner,ChiribellaValiron}.

Quantum Turing Machines \cite{Bernstein}, for instance, lie on this
other extreme, since the entire machine can be in a superposition of
base vectors. Unfortunately they are a rather oblivious way to
describe an algorithm.  Functional-style control structure, on the
other hand, seem to merge with quantum evolution descriptions in a
unifying manner. The functional language we describe may give rise to
a ``purely quantum'' programming language, but only once settled the
question of restricting to unitary operators.  This language is exempt
of classical registers, classical control structure, measurements, and
allows arbitrary quantum superpositions of base vectors.
 
A survey and comparison of these quantum programming languages can be found 
in \cite{Gay}.

\subsection{Current status of the language}
In our view, the problem of formulating a language of higher-order computable operators upon infinite dimensional vector spaces was
the first challenge that needed to be met, before even aiming to
have a physically executable language. In the current state of affairs
computability in vector spaces is dealt with matrices and
compositions, and hence restricted to finite-dimensional
systems -- although this limitation is sometimes circumvented by
introducing an extra classical control structure e.g. via the notions of
uniform circuits or linear types. The language we provide achieves
this goal of a minimal calculus for describing higher-order
computable linear operators in a wide sense.  Therefore this work may
serve as a basis for studying wider notions of computability upon
abstract vector spaces, whatever the interpretation of the vector space is (probabilities, number of computational paths leading to one result,\ldots).

The downside of this generality as far as the 
previously mentioned application to quantum computation are concerned is that our
operators are not restricted to being unitary. A further step towards
specializing our language to quantum computation would be to restrict
to unitary operators, as required by quantum physics.  There may be
several ways to do so. A first lead would be to design an a posteriori
static analysis that enforces unitarity -- exactly like typability is
not wired in pure lambda-calculus, but may be enforced a posteriori.
A second one would be to require a formal unitarity proof from the
programmer. With a term and a unitarity proof, we could derive a more
standard representation of the operator, for instance in terms of a
universal set of quantum gates \cite{Boykin}.  This transformation may be
seen as part of a compilation process. 

In its current state, our language can be seen as a specification language for quantum
programs, as it possesses several desirable features of such a
language: it allows a high level description of algorithms without any
commitment to a particular architecture, it allows the expression of
black-box algorithms through the use of higher-order functionals, its
notation remains close to both linear algebra and functional
languages.  

Besides quantum computing, this work may bring contributions to other
fields, which we now develop.

\VML{\smallskip}\noindent
\subsection{Logics, calculi}
In this article linearity is understood in the sense of linear
algebra, which is not to be confused with linearity in the sense of
Linear $\lambda$-calculus \cite{AbramskyLL}. It may help the reader to
draw attention to this distinction: Linear $\lambda$-calculus is a
calculus whose types are formulae of Linear Logic
\cite{Girard1}. \VML{In such a $\lambda$-calculus, one distinguishes
  linear resources, which may be neither duplicated nor discarded,
  from nonlinear ones, whose fate is not subjected to particular
  restrictions. The Linear-algebraic $\lambda$-calculus we describe
  does have some resemblances with the Linear $\lambda$-calculus, as
  well as some crucial, strongly motivated differences.  Duplication
  of a term ${\bf t}$ is again treated cautiously, but in a different
  way: only terms expressing base vectors can be duplicated, which is
  compatible with linear algebra. As we shall see, terms of the form
  $\lambda {\bf x}~{\bf u}$ are always base vectors. As a consequence,
  even when a term ${\bf t}$ cannot be duplicated the term $\lambda
  {\bf x}~{\bf t}$ can.  Since the term $\lambda {\bf x}~{\bf t}$ is a
  function building the term ${\bf t}$, it can be thought of as a
  description of ${\bf t}$. (This suggests some possible connections
  between the abstraction $\lambda {\bf x}$, the $!$ bang operator of
  linear lambda-calculus and the $'$ quote operator that transforms a
  term into a description of it as used for instance in
  LISP.)}\\ Again in connection with Linear Logic, Vaux has proposed
an {\em Algebraic $\lambda$-calculus} \cite{Vaux} independently and
simultaneously \cite{arrighidowek1,arrighidowek2,arrighidowek3} with
ours, and which is similar in the sense that it exhibits linear
combinations of terms, and different in both the reduction strategy
and the set of scalars considered. We will say more about this in
Section \ref{currentworks}. His work is both a restriction (less
operators) and a generalization (positive reals) of Ehrhard and
Regnier's differential $\lambda$-calculus \cite{Ehrhard}.\\ The
functional style of programming is based on the $\lambda$-calculus
together with a number of extensions, so as to make everyday
programming more accessible. Hence, since the birth of functional
programming there has been several theoretical studies of extensions
of the $\lambda$-calculus in order to account for basic arithmetic
(see for instance Dougherty's algebraic extension \cite{Dougherty} for
normalising terms of the $\lambda$-calculus). {\em Lineal} could again
be viewed as just an extension of the $\lambda$-calculus in order to
handle operations over vector spaces, and make everyday programming
more accessible upon them. The main difference in approach is that
here the $\lambda$-calculus is not seen as a control structure which
sits on top of the vector space data structure, controlling which
operations to apply and when. Rather, the $\lambda$-calculus terms
themselves can be summed and weighted, hence they actually are the
basis of the vector space\ldots upon which they can also act. This
intertwining of concepts is essential if seeking to represent parallel
or probabilistic computation as it is the computation itself which
must be endowed with a vector space structure. The ability to
superpose $\lambda$-calculus terms in that sense takes us back to
Boudol's parallel $\lambda$-calculus \cite{Boudol}, and may also be
viewed as taking part of a wave of probabilistic extensions of
calculi, e.g.\cite{Bournez,Catuscia,Hankin}.

\VML{\smallskip}\noindent
\subsection{Confluence techniques}\, A standard way to
describe how a program is executed is to give a small step operational
semantic for it, in the form of a finite set rewrite rules which 
gradually transform a program into a value. 
The main theorem proved in this paper is the confluence of our
language. What this means is that the order in which those
transformations are applied does not affect the end result of the
computation. Confluence results are milestones in the study of
programming languages and more generally in the theory of
rewriting. Our proof uses many of the theoretical tools that have been
developed for confluence proofs in a variety of fields (local
confluence and Newman's lemma; strong confluence and the 
Hindley-Rosen lemma) as well as the avatar lemma for parametric
rewriting as introduced in \cite{arrighidowek1}. 
These are fitted together in an elaborate architecture
which may have its own interest whenever one seeks to merge
a non-terminating conditional confluent rewrite system together with a terminating conditional confluent rewrite system.

\VML{\smallskip}\noindent
\subsection{Outline}
Section \ref{vecspaces} develops a computational definition of vector spaces and bilinear functions. This is achieved by taking the axioms of vector spaces and orienting them. Section \ref{fieldofqc} explains how to have a rewrite system for scalars that are enough to account for quantum computation. Section \ref{mainfeatures} presents the designing principles of the language, Section \ref{language} formally describes the Linear-algebraic $\lambda$-calculus and its semantics. Section \ref{encodings} shows that the language is expressive enough for classical and quantum computations. These are the more qualitative sections of the paper.
We chose to postpone till Section \ref{confluence} the various proofs of confluence, as they are more technical. Section \ref{currentworks} will be the opportunity to provide an overview of the most recent contributions surrounding this work. Section \ref{conclusion} provides a summary and some perspectives.

\section{Computational vector spaces and bilinear functions}\label{vecspaces}

One way to prove the equality of two vectors expressed by terms
such as $2 . {\bf x} + {\bf y} + 3 . {\bf x}$ and $5.({\bf x} +
{\bf y}) + (-4). {\bf y}$ is to transform these terms into linear
combinations of the unknowns and check that the terms obtained
this way are the same. This algorithm transforming a term
expressing a vector into a linear combination of the unknowns is
also useful to express the operational semantic of programming
languages for quantum computing, because in such
languages a program and its input value form a term expressing a
vector whose value, the output, is a linear combination of constants.
More generally, several algorithms used in linear algebra, such as
matrix multiplication algorithms, transform a term expressing a
vector with various constructs into a linear combination of
constants.

The algorithm transforming a term expressing a vector into a linear
combination of the unknowns is valid in all vector spaces.  The
goal of this Section is to show that, moreover, it completely
defines the notion of vector space. This computational definition
of the notion of vector space can be extended to define other
algebraic notions such as bilinearity.

\subsection{Algorithms and models}
\label{sectionmodels}

In this paper rewriting systems play a double role: they serve to provide an oriented version of the notion of vector space, and to provide a small step operational semantics for the introduced language. We now provide the standard definitions about them.

\begin{defi} {\bf (Rewriting)}
Let ${\cal L}$ be a first-order language. A rewrite system  $X$ on
${\cal L}$ is given by a finite set of rules of the form $l \lra r$. 
We define the relation $\lra_{X}$ as follows: $t \lra_{X} u$ if and only if  there is an occurrence $\alpha$ in the term $t$, a rewrite rule $l \lra r$ in $X$, and a substitution $\sigma$ such that $t_{|\alpha} = \sigma l$ and $u = t[\sigma  r]_{\alpha}$, where $t_{|\alpha}$ is the subterm of $t$ at occurrence $\alpha$, and $t[v]_{\alpha}$ is the graft of $v$ in $t$ occurrence $\alpha$.
\end{defi}

\begin{defi}{\bf (AC-Rewriting)}\label{defrewritingAC}
Let ${\cal L}$ be a first-order language. A AC-rewrite system $X$ on
${\cal L}$ is given by binary function symbols $f_1,\ldots,f_n$ of the language and a finite set of rules of the form $l \lra r$. 
We define the relation $=_{AC}$ as the congruence generated by the 
associativity and commutativity axioms of the symbols $f_1,\ldots,f_n$. 
We define the relation $\lra_{X}$
as follows: $t \lra_{X} u$ if and only if there exists a term $t'$ such that 
$t =_{AC} t'$, an occurrence $\alpha$ in $t'$, a rewrite rule $l \lra r$ in
$X$ and a substitution $\sigma$ such that $t'_{|\alpha} = \sigma l$ and $u =_{AC}
t'[\sigma r]_\alpha$.
\end{defi}

\begin{defi} {\bf (Algebra)}
Let ${\cal L}$ be a first-order language.  An {\em ${\cal
L}$-algebra} is a family formed by a set $M$ and for each symbol
$f$ of ${\cal L}$ of arity $n$, a function $\hat{f}$ from
$M^n$ to $M$. The denotation $\llbracket t \rrbracket_{\phi}$ of a term
$t$ for an assignment $\phi$ is defined as usual: $\llbracket x \rrbracket_{\phi}=\phi(x)$ and $\llbracket f(t_1,\ldots,t_n) \rrbracket_{\phi}=\hat{f}(\llbracket t_1,\ldots,t_n \rrbracket)$. 
\end{defi}

\begin{defi} {\bf (Model of a rewrite system)}
Let ${\cal L}$ be a first-order language and $X$ an algorithm
defined by a rewrite system on terms of the language ${\cal L}$.
An {\em ${\cal L}$-algebra} ${\cal M}$ is a {\em model} of the
algorithm $X$, or the algorithm $X$ is
{\em valid} in the model ${\cal M}$, (${\cal M} \models X$) if
for all rewrite rules
$l \lra r$ of the rewrite system and for all valuations $\phi$,
$\llbracket l \rrbracket_{\phi} =
\llbracket r \rrbracket_{\phi}$.
\end{defi}

\begin{exa}
Consider the language ${\cal L}$ formed by two binary symbols $+$ and
$\times$ and the algorithm $X$ defined by the distributivity rules
$$(x + y) \times z \lra (x \times z) + (y \times z)$$
$$x \times (y + z) \lra (x \times y) + (x \times z)$$
transforming for instance, the term $(a + a) \times a$ to
the term $a \times a + a \times a$.
The algebra $\langle \{0, 1\}, \mbox{\em min}, \mbox{\em
max} \rangle$
is a model of this algorithm.
\end{exa}

\begin{rem}
This definition of the validity of an algorithm in a model is strongly related with denotational semantics, as rewriting systems could also be seen as programs, and the algebraic structure as a denotational semantics. 
\end{rem}

\begin{defi} {\bf (Model of an AC-rewrite system)}
Let ${\cal L}$ be a first-order language. Let $X$ be a AC-rewrite system.
An {\em ${\cal L}$-algebra} ${\cal M}$ is a {\em model}
of the AC-rewrite system $X$ (${\cal M} \models R$) if
\begin{itemize}
\item
for all rewrite rules
$l \lra r$ of $X$ and for all valuations $\phi$,
$\llbracket l \rrbracket_{\phi} =
\llbracket r \rrbracket_{\phi}$,
\item for all AC-symbol $f$ of $X$ and
for all valuations $\phi$ and indices $i$
$$\llbracket f(x, f(y,z)) \rrbracket_{\phi} = \llbracket
f(f(x, y),  z) \rrbracket_{\phi}$$
$$\llbracket f(x, y) \rrbracket_{\phi} = \llbracket f(y,x)
\rrbracket_{\phi}$$
\end{itemize}
\end{defi}\bigskip

\noindent As a consequence if $t\lra_{X}^*u$ then for all $\phi$, $\llbracket t \rrbracket_\phi=\llbracket u \rrbracket_\phi$.

\subsection{Vector spaces: an algorithm}
\label{sectionvecspaces}

Let ${\cal L}$ be a 2-sorted language with a sort $K$ for scalars and
a sort $E$ for vectors containing
two binary symbols $+$ and $\times$ of rank $\langle K, K, K \rangle$,
two constants $0$ and $1$ of sort $K$,
a binary symbol, also written $+$, of rank
$\langle E, E, E \rangle$, a binary symbol $.$ of rank
$\langle K, E, E \rangle$ and a constant ${\bf 0}$ of sort $E$.

To transform a term of sort $E$ into a linear combination of the unknowns,
we want to develop sums of vectors: $\alpha . ({\bf u} + {\bf v}) \lra \alpha . {\bf u} + \alpha . {\bf v}$, but factor sums of scalars and nested products: $\alpha . {\bf u} + \beta . {\bf u} \lra (\alpha + \beta) . {\bf u}$, $\alpha . (\beta . {\bf u}) \lra (\alpha \times \beta) . {\bf u}$. We also need the trivial rules ${\bf u} + {\bf 0}  \lra {\bf u}$, $0 . {\bf u} \lra {\bf 0}$ and $1 . {\bf u} \lra {\bf u}$. Finally, we need three more rules for confluence 
$\alpha . {\bf 0} \lra {\bf 0}$, $\alpha . {\bf u} + {\bf u} \lra (\alpha + 1) . {\bf u}$, ${\bf u} + {\bf u} \lra (1 + 1) . {\bf u}$. As we want to be able to apply the factorization rule to a term of the form $(3.{\bf x} + 4.{\bf y}) + 2.{\bf x}$, reductions in the above rewrite system must be defined modulo the associativity and commutativity of $+$. This leads to the following definition.

\begin{defi}\label{def:V} {\bf (The rewrite system $V$)}
The rewrite system $V$ is the AC-rewrite system
where the only AC-symbol is $+$ and the rules are 
$${\bf u} + {\bf 0}  \lra {\bf u}$$
$$0 . {\bf u} \lra {\bf 0}$$
$$1 . {\bf u} \lra {\bf u}$$
$$\alpha . {\bf 0} \lra {\bf 0}$$
$$\alpha . (\beta . {\bf u}) \lra (\alpha . \beta). {\bf u}$$
$$\alpha . {\bf u} + \beta . {\bf u} \lra (\alpha + \beta) . {\bf u}$$
$$\alpha . {\bf u} + {\bf u} \lra (\alpha + 1) . {\bf u}$$
$${\bf u} + {\bf u} \lra (1 + 1) . {\bf u}$$
$$\alpha . ({\bf u} + {\bf v}) \lra \alpha . {\bf u} + \alpha . {\bf v}$$
\end{defi}

To be complete, we should also transform the axioms of the theory of
fields into a rewrite system, which is known to be impossible as there
is no equational description of fields as a consequence of Birkhoff's HSP
theorem and the fact that the class of fields is not closed under
direct product \cite{Birkhoff}.  

We could switch to the theory of modules and use the fact that 
the axioms of the theory of rings can be transformed into a rewrite 
system. 

An alternative is to provide term rewrite systems for specific rings
or fields such as ${\mathbb Z}$, ${\mathbb Q}$, ${\mathbb
  Q}(i,\sqrt{2})$, etc.  as we shall do in Section \ref{fieldofqc}.
Notice that these rewrite system are in general richer than that of
the theory of rings. For instance in the language of the rewrite
system of ${\mathbb Q}(i,\sqrt{2})$, we have terms expressing the
numbers $1/2$ or $\sqrt{2}$ that are not in the generic language of
rings.

Thus we shall introduce a general notion of ``scalar rewrite system''
and consider an arbitrary such system.  Basically the notion of a
scalar rewrite systems lists the few basic properties that scalars are
usually expected to have: neutral elements, associativity of $+$, etc.

\begin{defi}\label{scalarrs} {\bf (Scalar rewrite system)}
A {\em scalar rewrite system} is a
rewrite system on a language containing at least the symbols $+$,
$\times$, $0$ and $1$ such that:

\begin{itemize}
\item $S$ is terminating and ground confluent,

\item for all closed terms
$\alpha$, $\beta$ and $\gamma$, the pair of terms
\begin{itemize}
\item $0 + \alpha$ and $\alpha$,
\item $0 \times \alpha$ and $0$,
\item $1 \times \alpha$ and $\alpha$,
\item $\alpha \times (\beta + \gamma)$ and $(\alpha \times \beta) +
(\alpha \times \gamma)$,
\item $(\alpha + \beta) + \gamma$ and $\alpha + (\beta + \gamma)$,
\item $\alpha + \beta$ and $\beta + \alpha$,
\item $(\alpha \times \beta) \times \gamma$ and $\alpha \times (\beta \times \gamma)$,
\item $\alpha \times \beta$ and $\beta \times \alpha$
\end{itemize}
have the same normal forms,

\item $0$ and $1$ are normal terms.
\end{itemize}
\end{defi}
Later in Subsection \ref{appcp1} we shall prove that for any such scalar rewrite system $S$, $S\cup V$ is terminating and confluent.

\begin{prop}
\label{classification}
Let ${\bf t}$ be a normal term whose variables are among
${\bf x}_{1}, ..., {\bf x}_{n}$.
The term ${\bf t}$ is ${\bf 0}$ or a term of the form
$\alpha_{1} . {\bf x}_{i_{1}} + ... +
\alpha_{k} . {\bf x}_{i_{k}} +
{\bf x}_{i_{k+1}} + ... +
{\bf x}_{i_{k+l}}$
where the indices $i_{1}, ..., i_{k+l}$ are distinct
and $\alpha_{1}, ..., \alpha_{k}$ are neither $0$ nor $1$.

\end{prop}

\begin{proof}
The term ${\bf t}$ is a sum ${\bf u}_{1} + ... + {\bf u}_{n}$ of
normal terms that are not sums (we take $n = 1$ if ${\bf t}$ is not a
sum).

A normal term that is not a sum is either ${\bf 0}$,
a variable, or a term of the form $\alpha . {\bf v}$. In this case,
$\alpha$ is neither $0$ nor $1$ and
${\bf v}$ is neither ${\bf 0}$, nor a sum of two vectors
nor a product of a scalar by a vector, thus it is a variable.

As the term ${\bf t}$ is normal, if $n > 1$ then none of the ${\bf
u}_{i}$ is ${\bf 0}$. Hence, the term ${\bf t}$ is either ${\bf 0}$
or a term of the form

$$\alpha_{1} . {\bf x}_{i_{1}} + ... + \alpha_{k} . {\bf x}_{i_{k}} +
{\bf x}_{i_{k+1}} + ... + {\bf x}_{i_{k+l}}$$
where $\alpha_{1}, ..., \alpha_{k}$ are neither $0$ nor $1$.
As the term {\bf t} is normal, the indices $i_{1}, ..., i_{k+l}$ are distinct.
\end{proof}

\subsection{Vector spaces: a computational characterization} 

With respect to the notion of model, algorithms play the same role as
sets of axioms: an algorithm may or may not be valid in a model, exactly
like a set of axioms may or may not be valid in a model.

The notion of validity may be used to study sets of axioms,
typically building a model is a way to prove that some proposition
is not provable from a set of axioms.
But validity can also be used in the other direction: to define algebraic structures as models of some theories. For
instance, given a field
${\cal K} = \langle K, +, \times, 0, 1 \rangle$
the class of ${\cal K}$-vector spaces can be defined as follows.

\begin{defi} {\bf (Vector space)}
\label{def1}
The algebra $\langle E, +, .,
{\bf 0} \rangle$ is a ${\cal K}$-vector space if and only if ${\cal K}=\langle K, +, \times, 0,1 \rangle$ is a field and the
algebra $\langle K, +, \times, 0, 1,
E, +, ., {\bf 0} \rangle$ is a model of the 2-sorted
set of axioms

$$\forall {\bf u} \forall {\bf v} \forall {\bf w}~(
({\bf u} + {\bf v}) + {\bf w} = {\bf u} + ({\bf v} + {\bf w}))$$
$$\forall {\bf u} \forall {\bf v}~(
{\bf u} + {\bf v}  = {\bf v} + {\bf u})$$
$$\forall {\bf u}~(
{\bf u} + {\bf 0}  = {\bf u})$$
$$\forall {\bf u}~\exists {\bf u'}~({\bf u} + {\bf u'} = {\bf 0})$$
$$\forall {\bf u}~
(1. {\bf u} = {\bf u})$$
$$\forall \alpha \forall \beta \forall {\bf u}~
(\alpha . (\beta. {\bf u}) = (\alpha . \beta). {\bf u})$$
$$\forall \alpha \forall \beta \forall {\bf u}~
((\alpha + \beta) . {\bf u} = \alpha . {\bf u} + \beta . {\bf u})$$
$$\forall \alpha \forall {\bf u} \forall {\bf v}~
(\alpha . ({\bf u} + {\bf v}) = \alpha . {\bf u} + \alpha . {\bf v})$$
\end{defi}\bigskip

\noindent We now prove that, the class of ${\cal K}$-vector spaces can be
defined as the class of models of the rewrite system $V$.

\begin{prop}\label{prop:evequivalence}
Let ${\cal K} = \langle K, +, \times, 0, 1 \rangle$
be a field. The algebra $\langle E, +, .,
{\bf 0} \rangle$ is a ${\cal K}$-vector space if and only if the
algebra $\langle K, +, \times, 0, 1,
E, +, ., {\bf 0} \rangle$ is a model of the
rewrite system $V$.
\end{prop}

\begin{proof}
We first check that all the rules of $V$ and the associativity and commutativity of addition are valid in all vector spaces. All of them are trivial except 
$\alpha . {\bf u} + {\bf u} = (\alpha + 1) . {\bf u}$,
${\bf u} + {\bf u} =  (1 + 1) . {\bf u}$,
$0. {\bf u} = {\bf 0}$ and
$\alpha . {\bf 0} = {\bf 0}$.
The first and second are consequence of $1 . {\bf u} = {\bf u}$ and
$\alpha . {\bf u} + \beta . {\bf u} = (\alpha + \beta) . {\bf u}$.
To prove the third let ${\bf u'}$ be such that  ${\bf u} + {\bf u'}  = {\bf 0}$. We have
$0 . {\bf u} = 0 . {\bf u} + {\bf 0}
= 0 . {\bf u} + {\bf u} + {\bf u'}
= 0 . {\bf u} + 1 . {\bf u} + {\bf u'}
= 1 . {\bf u} + {\bf u'}
= {\bf u} + {\bf u'} = {\bf 0}$.
The last one is a consequence of
$0. {\bf u} = {\bf 0}$ and
$\alpha . (\beta . {\bf u}) = (\alpha. \beta). {\bf u}$.

Conversely, we prove that all axioms of vector spaces are
valid in all models of $V$. The validity of each of them is a consequence
of the validity of a rewrite rule, except
$\forall {\bf u} \exists {\bf u'}~({\bf u} + {\bf u'}  = {\bf 0})$
that is a consequence of
${\bf u} + (-1) . {\bf u} = {\bf 0}$
itself being a consequence of
$\alpha . {\bf u} + \beta . {\bf u} = (\alpha + \beta) . {\bf u}$
and $0 . {\bf u} = {\bf 0}$.
\end{proof}

\subsection{Vector spaces: decidability}

We now show that the rewrite system $V$ (Definition \ref{def:V}) permits to prove the decidability of the word problem (i.e., whether two terms express the same vector or not) for vector spaces. 

\begin{defi}
The {\em decomposition} of ${\bf t}$ along
${\bf x}_{1}, ..., {\bf x}_{n}$
is the sequence
$\alpha_{1}, ..., \alpha_{n}$
such that if there is a subterm of the form
$\alpha . {\bf x}_{i}$ in ${\bf t}$,
then $\alpha_{i} = \alpha$,
if there is a subterm of the form
${\bf x}_{i}$ in ${\bf t}$,
then $\alpha_{i} = 1$, and $\alpha_{i} = 0$ otherwise.
\end{defi}

\begin{prop}\label{prop:evdecidability}
Let ${\bf t}$ and ${\bf u}$ be two terms whose variables are among
${\bf x}_{1}, ..., {\bf x}_{n}$. The following propositions are equivalent:
\begin{enumerate}[label=(\roman*)]
\item[(i)] the normal forms of ${\bf t}$ and ${\bf u}$ are identical modulo AC,
\item[(ii)] the equation ${\bf t} = {\bf u}$ is valid in all ${\cal
K}$-vector spaces,
\item[(iii)] and the denotation of ${\bf t}$ and ${\bf u}$ in $K^n$ for the
assignment $\phi = {\bf e}_{1} / {\bf x}_{1}, ..., {\bf e}_{n} / {\bf
  x_{n}}$,
where
${\bf e}_{1}, ..., {\bf e}_{n}$ is the canonical base of $K^n$,
are identical.
\end{enumerate}
\end{prop}

\begin{proof}
Proposition (i) implies proposition (ii)
and proposition (ii) implies proposition (iii). Let us prove that
proposition (iii) implies proposition (i).

Let ${\bf t}$ be a normal term
whose variables are among ${\bf x}_{1}, ..., {\bf x}_{n}$.
Assume $\llbracket {\bf t} \rrbracket_{\phi} = \llbracket {\bf u}
\rrbracket_{\phi}$.
Let
${\bf e}_{1}, ..., {\bf e}_{n}$ be the canonical base of
$K^n$ and $\phi = {\bf e}_{1}/{\bf x}_{1}, ..., {\bf e}_{n}/{\bf x}_{n}$.
Call $\alpha_{1}, ..., \alpha_{n}$ the coordinates of
$\llbracket {\bf t} \rrbracket_{\phi}$
in ${\bf e}_{1}, ..., {\bf e}_{n}$.
Then the decompositions of the normal forms of ${\bf t}$ and ${\bf u}$ are
both $\alpha_{1}, ..., \alpha_{n}$ and thus they are
identical modulo AC.
\end{proof}

\subsection{Summary}

We usually define an algebraic structure as an algebra $\langle M, \hat{f}_1,\ldots, \hat{f}_n\rangle $ that validates some propositions.
For instance ${\cal K}$-vector spaces are defined
as the algebras $\langle E, \mathbf{0}, +, .\rangle$ that validate the equations
of Definition \ref{def1}.

We can, in a more computation-oriented way, define an algebraic structure as an algebra that validates an algorithm on terms constructed upon
these operations.
For instance ${\cal K}$-vector spaces are are defined
as the algebras $\langle E, \mathbf{0}, +, .\rangle$ that validate the
algorithm $V$ of Definition \ref{def:V}.

This algorithm is a well-known algorithm in linear algebra: it is the
algorithm that transforms any linear expression into a linear
combination of the unknowns. If we chose a base, as will be the case in 
section \ref{language}, this algorithm may be used to transforms any linear 
expression into a linear combination of base vectors. 
Still the algorithm itself is not linked to any particular base and it 
may even be used if the unknowns represent a linearly dependent family.

This algorithm is, at a first look,
only one among the many algorithms used in linear algebra,
but it completely
defines the notion of vector space: a vector space is any
algebra where this algorithm is valid, it is any algebra
where linear expressions can be transformed this way into linear
combinations of the unknowns.

\subsection{Bilinearity}
\label{sectiontensors}

Another important notion about vector spaces is that of bilinear functions. For instance the tensor product, matrix multiplication, the inner product and as we shall see the application in Lineal are all bilinear operations. The method we developed for a computational characterization of vector spaces extends to this notion:
\begin{defi} {\bf (Bilinear function)}
\label{def1'}
Let $E$, $F$, and $G$ be three vector spaces on the same
field. A function $\otimes$ from $E \times F$ to $G$ is said to be
{\em bilinear} if

$$({\bf u} + {\bf v}) \otimes {\bf w} =
({\bf u} \otimes {\bf w}) + ({\bf v} \otimes {\bf w})$$
$$(\alpha . {\bf u}) \otimes {\bf v} =
\alpha . ({\bf u} \otimes {\bf v})$$
$${\bf u} \otimes ({\bf v} + {\bf w}) =
({\bf u} \otimes {\bf v}) + ({\bf u} \otimes {\bf w})$$
$${\bf u} \otimes (\alpha . {\bf v}) =
\alpha . ({\bf u} \otimes {\bf v})$$
\end{defi}

\begin{defi} {\bf (Tensor product)}
Let $E$ and $F$ be two vector spaces, the pair formed by the vector
space $G$ and the bilinear function from $E \times F$ to $G$ is a
{\em tensor product} of $E$ and $F$ if for all bases
$({\bf e}_{i})_{i \in I}$ of $E$ and $({\bf e'}_{j})_{j \in J}$ of $F$ the
family $({\bf e}_{i} \otimes {\bf e'}_{j})_{\langle i,j \rangle}$ is a
base of $G$.
\end{defi}

The corresponding algorithm is as follows:
\begin{defi} {\bf (The rewrite system $V'$)}
\label{V'}
Consider a language with four sorts: $K$ for scalars and $E$, $F$,
and $G$ for the vectors of three vector spaces, the symbols $+$, $\times$,
$0$, $1$ for scalars, three copies of the symbols $+$, $.$ and ${\bf 0}$ for
each sort $E$, $F$, and $G$ and a symbol $\otimes$ of rank
$\langle E, F, G \rangle$.

The system $V'$ is the rewrite system formed by three copies of
the rules of the system $V$ and the rules

$$({\bf u} + {\bf v}) \otimes {\bf w} \lra
({\bf u} \otimes {\bf w}) + ({\bf v} \otimes {\bf w})$$
$$(\alpha . {\bf u}) \otimes {\bf v} \lra
\alpha . ({\bf u} \otimes {\bf v})$$
$${\bf u} \otimes ({\bf v} + {\bf w}) \lra
({\bf u} \otimes {\bf v}) + ({\bf u} \otimes {\bf w})$$
$${\bf u} \otimes (\alpha . {\bf v}) \lra
\alpha . ({\bf u} \otimes {\bf v})$$
$${\bf 0} \otimes {\bf u} \lra {\bf 0}$$
$${\bf u} \otimes {\bf 0} \lra {\bf 0}$$
\end{defi}\bigskip

\noindent Later in Subsection \ref{appcp1} we shall prove that for any such scalar rewrite system $S$, $S\cup V'$ is terminating and confluent.

Propositions \ref{classification}-\ref{prop:evdecidability} generalize easily. 
\begin{prop}
Let ${\bf t}$ be a normal term whose variables of sort $E$ are among
${\bf x}_{1}, ..., {\bf x}_{n}$, whose variables of sort $F$ are
among ${\bf y}_{1}, ..., {\bf y}_{p}$, and that has
no variables of sort $G$ and $K$.
If ${\bf t}$ has sort
$E$ or $F$, then it has the same form as in Proposition
\ref{classification}. If it has sort $G$, then
it has the form

$$\alpha_{1} . ({\bf x}_{i_{1}} \otimes {\bf y}_{j_{1}})
+ ... +
\alpha_{k} . ({\bf x}_{i_{k}} \otimes {\bf y}_{j_{k}})
+
({\bf x}_{i_{k+1}} \otimes {\bf y}_{j_{k+1}})
+ ... +
({\bf x}_{i_{k+l}} \otimes {\bf y}_{j_{k+l}})$$
where the pairs of indices $\langle i_{1},j_{1} \rangle, ...,
\langle i_{k+l}, j_{k+l} \rangle$ are distinct
and $\alpha_{1}, ..., \alpha_{k}$ are neither $0$ nor $1$.
\end{prop}
\couic{
\begin{proof}
The term ${\bf t}$ is a sum ${\bf u}_{1} + ... + {\bf u}_{n}$ of
normal terms that are not sums (we take $n = 1$ if ${\bf t}$ is not a
sum).

A normal term that is not a sum is either ${\bf 0}$,
a term of the form ${\bf v} \otimes {\bf w}$,
or of the form $\alpha . {\bf v}$. In this case,
$\alpha$ is neither $0$ nor $1$ and
${\bf v}$ is neither ${\bf 0}$, nor a sum of two vectors
nor a product of a scalar by a vector, thus it is of the form
${\bf v} \otimes {\bf w}$.

In a term of the form ${\bf v} \otimes {\bf w}$, neither
${\bf v}$ nor ${\bf w}$ is a sum, a product of a scalar by a vector
or ${\bf 0}$. Thus both ${\bf v}$ and ${\bf w}$ are variables.

As the term ${\bf t}$ is normal, if $n > 1$ then none of the ${\bf
u}_{i}$ is ${\bf 0}$. Hence, the term ${\bf t}$ is either ${\bf 0}$
or a term of the form
$\alpha_{1} . ({\bf x}_{i_{1}} \otimes {\bf y}_{j_{1}})
+ ... +
\alpha_{k} . ({\bf x}_{i_{k}} \otimes {\bf y}_{j_{k}})
+
({\bf x}_{i_{k+1}} \otimes {\bf y}_{j_{k+1}})
+ ... +
({\bf x}_{i_{k+l}} \otimes {\bf y}_{j_{k+l}})$
where $\alpha_{1}, ..., \alpha_{k}$ are neither $0$ nor $1$.
As the term {\bf t} is normal, the pairs of indices are distinct.
\end{proof}
}
\couic{
\subsection{Bilinearity: a computational characterization}
}

\begin{prop}
Let ${\cal K} = \langle K, +, \times, 0, 1 \rangle$ be a field.
The structures
$\langle E, +, ., {\bf 0} \rangle$,
$\langle F, +, ., {\bf 0} \rangle$,
$\langle G, +, ., {\bf 0} \rangle$ are ${\cal K}$-vector spaces and
$\otimes$ is a bilinear function from $E \times F$ to $G$ if and
only if
$\langle K, +, \times, 0, 1,
E, +, ., {\bf 0},
F, +, ., {\bf 0},
G, +, ., {\bf 0}, \otimes \rangle$ is a model of the system $V'$.
\end{prop}
\couic{
\begin{proof}
The validity of the rules of the three copies of the system $V$, express
that
$\langle E, +, ., {\bf 0} \rangle$,
$\langle F, +, ., {\bf 0} \rangle$,
$\langle G, +, ., {\bf 0} \rangle$ are ${\cal K}$-vector spaces.
The validity of the six other rules is the validity of the axioms of
Definition \ref{def1'} plus the two extra propositions
${\bf 0} \otimes {\bf u} = {\bf 0}$ and
${\bf u} \otimes {\bf 0} = {\bf 0}$
that are consequences of these axioms.
\end{proof}
}

\couic{\subsection{Bilinearity: decidability}}

\begin{prop}
Let ${\bf t}$ and ${\bf u}$ be two terms whose variables of sort $E$ are among
${\bf x}_{1}, ..., {\bf x}_{n}$, whose variables of sort $F$ are
among ${\bf y}_{1}, ..., {\bf y}_{p}$, and that have
no variables of sort $G$ and $K$. The following propositions are equivalent:
\begin{enumerate}[label=(\roman*)]
\item[(i)] the normal forms of ${\bf t}$ and ${\bf u}$ are identical modulo AC,
\item[(ii)] the equation ${\bf t} = {\bf u}$ is valid in all structures formed
by three vector spaces and a bilinear function,
\item[(iii)] the equation ${\bf t} = {\bf u}$ is valid in all structures formed
by two vector spaces and their tensor product,
\item[(iv)] and the denotation of ${\bf t}$ and ${\bf u}$ in $K^{np}$ for the
assignment

$$\phi = {\bf e}_{1} / {\bf x}_{1}, ..., {\bf e}_{n} / {\bf x_{n}},
{\bf e'}_{1} / {\bf y}_{1}, ..., {\bf e'}_{p} / {\bf y_{p}}$$
where
${\bf e}_{1}, ..., {\bf e}_{n}$ is the canonical base of $K^n$,
${\bf e'}_{1}, ..., {\bf e'}_{p}$ that of $K^p$
and $\otimes$ is the
unique bilinear function such
that
${\bf e}_{i} \otimes {\bf e'}_{j} = {\bf e''}_{p(i-1)+j}$
where
${\bf e''}_{1}, ..., {\bf e''}_{np}$ is the canonical base of $K^{np}$.
\end{enumerate}
\end{prop}
\couic{
\begin{proof}
Proposition (i) implies proposition (ii),
proposition (ii) implies proposition (iii) and
proposition (iii) implies proposition (iv).
Let us prove that proposition (iv) implies proposition (i).

Let ${\bf t}$ be a normal term of sort $G$ with variables
of sort $E$ among ${\bf x}_{1}, ..., {\bf x}_{n}$, variables
of sort $F$ among
${\bf y}_{1}, ..., {\bf y}_{p}$, and no variables of sort $G$ and $K$.
The {\em decomposition} of ${\bf t}$ along
${\bf x}_{1}, ..., {\bf x}_{n}$,
${\bf y}_{1}, ..., {\bf y}_{p}$,
is the sequence
$\alpha_{1}, ..., \alpha_{np}$
such that if there is a subterm of the form
$\alpha . ({\bf x}_{i} \otimes {\bf y}_{j})$ in ${\bf t}$,
then $\alpha_{p (i - 1) +j} = \alpha$,
if there is a subterm of the form
${\bf x}_{i} \otimes {\bf y}_{j}$ in ${\bf t}$,
then $\alpha_{p (i - 1) + j} = 1$, and $\alpha_{p (i - 1) + j} = 0$ otherwise.

Assume $\llbracket {\bf t} \rrbracket_{\phi} = \llbracket {\bf u}
\rrbracket_{\phi}$.
Call $\alpha_{1}, ..., \alpha_{np}$ the coordinates of
$\llbracket {\bf t} \rrbracket_{\phi}$
in ${\bf e''}_{1}, ..., {\bf e''}_{np}$.
Then the decompositions of the normal forms of ${\bf t}$ and ${\bf u}$ are
both $\alpha_{1}, ..., \alpha_{np}$ and thus they are
identical modulo AC.
\end{proof}

Again notice that in the above Proposition, $(i)\Leftrightarrow (ii)$ is a statement of the decidability of equality, whereas $(ii)\Leftrightarrow (iii)\Leftrightarrow (iv)$ is a statement of the universality of this notion of equality.
}

\section{The field of quantum computing}\label{fieldofqc}

As explained in Section \ref{sectionvecspaces}, 
fields are not easily implemented as term rewrite systems.
In the previous
section such problems were avoided by simply assuming the provision of
{\em some} scalar rewrite system, i.e., some term rewrite system for
scalars having a certain number of properties (Definition
\ref{scalarrs}).  However if the objective is to provide a formal
operational semantics for a quantum programming language, up to the
point that it provides a full description of a classical simulator for
the language, then we must give such a term rewrite system
explicitly. The present section briefly outlines how this can be
achieved. 

\subsection{A rewrite system for the field ${\mathbb Q}(i,\sqrt{2})$}
\label{subsecfield}

In the circuit model of quantum computation the emphasis was
placed on the ability to \emph{approximate} any unitary transform
from a finite set of gates, where approximation is defined in terms 
of the distance induced by the supremum norm.
This line of research (cf.
\cite{Solovay1,Kitaev} to cite a few) has culminated
with \cite{Boykin}, where the following set
\begin{align}
CNOT&=\left(\begin{array}{cccc}
1    &0    &0   &0\\
0    &1    &0   &0\\
0    &0    &0   &1\\
0    &0    &1   &0 \end{array}\right)\label{basicgates}\\
H&=\left(\begin{array}{cc}
\frac{1}{\sqrt{2}}     &\frac{1}{\sqrt{2}} \\
\frac{1}{\sqrt{2}}     &-\frac{1}{\sqrt{2}}
\end{array}\right)\quad
P=\left(\begin{array}{cc}
1     &0 \\
0     &e^{i\pi/4}
\end{array}\right)\nonumber
\end{align}
was proven to be universal, in the sense that its closure under composition, tensor product and tracing out forms a dense set relative to the set of unitary matrices --- with respect to the supremum norm.
Thus, the field $\mathbb{Q}(i,\sqrt{2})$ is enough for quantum computation.

To define a scalar rewrite system for this field we can proceed in three steps.
\begin{itemize}
\item We define a scalar rewrite system for ${\mathbb Q}$, 

\item We use Section \ref{sectionvecspaces}
to define a rewrite system for the vector space
of linear combination of the form:
\begin{equation*}
\alpha . \mathbf{1} + \beta . \mathbf{\frac{1}{\sqrt{2}}} + \gamma
. \mathbf{i} + \delta . \mathbf{\frac{i}{\sqrt{2}}}.
\end{equation*}

\item We use Section \ref{sectiontensors} to 
define multiplication as a bilinear AC-operation and with the rules 
\begin{align*}
\mathbf{1} \times \mathbf{v} &\lra \mathbf{v}
&\quad 
\mathbf{\frac{1}{\sqrt{2}}}  \times  \mathbf{\frac{1}{\sqrt{2}}} &\lra
(1/2).\mathbf{1}\\
\mathbf{\frac{1}{\sqrt{2}}} \times  \mathbf{i} &\lra \mathbf{\frac{i}{\sqrt{2}}}
&\quad 
\mathbf{\frac{1}{\sqrt{2}}} \times  \mathbf{\frac{i}{\sqrt{2}}} &\lra
(1/2).\mathbf{i}\\
\mathbf{i}  \times  \mathbf{i} &\lra 
(-1/2).\mathbf{1}
&\quad 
\mathbf{i}  \times  \mathbf{\frac{i}{\sqrt{2}}} &\lra 
(-1/2). \mathbf{\frac{1}{\sqrt{2}}}\\
\mathbf{\frac{i}{\sqrt{2}}} \times \mathbf{\frac{i}{\sqrt{2}}} &\lra
(-1/2). \mathbf{1}
\end{align*}
\end{itemize}

\subsection{Restricting to the ring ${\mathbb D}(i,1/\sqrt{2})$}

For the sake of implementation we can make further simplifications.
As division does not appear when composing unitary operators, 
the only scalars that appear in the closure of gates 
(\ref{basicgates})
are those 
obtained by additive and multiplicative closure from the
elements $1$, $i$ and $ \frac{1}{\sqrt{2}}$, i.e., elements of the ring
${\mathbb D}(i,1/\sqrt{2})$ where ${\mathbb D}$ is the ring of
numbers that have a finite dyadic development. Thus this ring is enough for 
quantum computing.

Notice that, as noticed above, switching to the theory of ring would not
be sufficient as, in order to express the gates of quantum computing, we 
need terms
expressing the scalars $i$ and $1/\sqrt{2}$. 
An implementation of this ring, along these lines, can be found in
\cite{arrighidowek4} that builds upon Section \ref{subsecfield} and
\cite{Cohen,Walters} for implementing binary numbers.

\section{Towards a higher-order language}\label{mainfeatures}

We introduce a language combining higher-order computation and linear
algebra. The syntax of this language is minimal, in the sense that it
just contains the syntax of $\lambda$-calculus and the possibility to
make linear combinations of terms $\alpha.{\bf t}+\beta.{\bf u}$. This
language is called Linear-algebraic $\lambda$-calculus, or just {\em
Lineal}.

To start with, we take as the operational semantics, the rewrite rules 
of vector spaces of Section \ref{sectionvecspaces}, the bilinearity of
application of Section \ref{sectiontensors}, and the $\beta$-reduction rule. 
The goal of this section is to present the issues arising. This will 
result in the fine-tuned system presented in Definition \ref{defsystemL}.

\VML{\smallskip}\noindent
\subsection{Higher-order $\lambda$-calculus}
In quantum computing, many algorithms fall into
the category of {\em black-box} algorithms. I.e., some mysterious
implementation of a function ${f}$ is provided to us which we call
``oracle'' -- and we wish to evaluate some property of ${f}$, after
a limited number of queries to its oracle. For instance in the
Deutsch-Josza quantum algorithm, $f$ is a function
$f:\{{\bf false},{\bf true}\}^n\lra \{{\bf false},{\bf true}\}$ which
is either constant (i.e., $\exists c\forall x[f(x) = c]$) or balanced
(i.e.,  $\vert\{x\,\textrm{such that}\,f(x)=
{\bf false}\}\vert=\vert\{x\,\textrm{such that}\,f(x)=
{\bf true}\}\vert$), and whose corresponding oracle is a unitary
transformation $U_f:\mathcal{H}_2^{n+1}\lra\mathcal{H}_2^{n+1}$ such
that $U_f: {\bf x} \otimes {\bf b}\mapsto {\bf x} \otimes ({\bf b} \oplus
f({\bf x}))$, where $\mathcal{H}_2^{n+1}$ stands for a tensor product of
$n+1$ two-dimensional vector spaces, $\otimes$ is the tensor product and $\oplus$
just the addition modulo two. The aim is to determine whether $f$ is constant
or balanced, and it turns out that this can be done in one single query to its
oracle. The algorithm works by applying $H^{\otimes^{n+1}}$ upon
$({\bf false}^{\otimes^{n}} \otimes {\bf true})$, then $U_f$, and then
$H^{\otimes^{n+1}}$ again, where $H^{\otimes^{n+1}}$ means applying the Hadamard gate
on each of the $n+1$ qubits.  It is clear that a
desirable feature for a linear-algebraic functional language 
is to be able to express algorithms as a
function of an oracle. E.g. we may want to define
$$\textbf{Dj}_1\equiv \lambda {\bf x}\, ((H\otimes
H)\,({\bf x}\,((H\otimes H)\, (\textbf{false}\otimes\textbf{true})))$$
so that $\textbf{Dj}_1\,U_f$ reduces to $(H\otimes
H)\,(U_f\,((H\otimes H)\,(\textbf{false}\otimes\textbf{true})))$.
More importantly even, one must be able to express algorithms, whether
they are black-box or not, independent of the size of their
input. This is what differentiates programs from fixed-size circuits
acting upon finite dimensional vector spaces,
and demonstrates the ability to have control flow. The way to achieve
this in functional languages involves duplicating basic components of
the algorithm an appropriate number of times.  E.g. we may want to
define some ${\bf Dj}$ operator so that $(\textbf{Dj}~{\bf n})~U_f$
reduces to the appropriate $(\textbf{Dj}_n)~U_f$, where ${\bf n}$ is a
natural number.

Clearly the language of uniform circuits does not offer an elegant
presentation for this issue.
Higher-order appears to be a desirable
feature to have for black-box computing, but also for expressing
recursion and for high-level programming.

\VML{\smallskip}\noindent
\subsection{Reduction strategies: copying and cloning}

We seek to design a $\lambda$-calculus, i.e., have
the possibility to introduce and abstract upon variables, as a mean to
express functions of these variables. In doing so, we must allow
functions such as $\lambda {\bf x}\, ({\bf x}\otimes {\bf x})$,
which duplicate their argument. This is necessary for expressiveness,
for instance in order to obtain the fixed point operator or any other
form of iteration/recursion.

Now problems come up when functions
such as $\lambda {\bf x}\, ({\bf x}\otimes {\bf x})$ are applied to 
superpositions (i.e., sums of vectors). 
Linear-algebra brings a strong constraint: we know that cloning
is not allowed, i.e., that the operator which maps any vector $\psi$ onto the
vector $\psi\otimes\psi$ is not linear. 
In quantum computation this impossibility is referred to as the {\em no-cloning} theorem \cite{cloning}.
Most quantum programming language proposals so far consist in some quantum
registers undergoing unitary transforms and measures on the one hand,
together with classical registers and programming structures ensuring
control flow on the other, precisely in order to avoid such problems. 
But as we seek to reach beyond this duality and obtain a purely 
quantum programming language, we need to face it in a different manner.

This problem may be seen as a confluence problem.  Faced with the term
$(\lambda {\bf x}\, ({\bf x} \otimes {\bf x})) \, ({\bf false}+{\bf
  true})$, one could either start by substituting ${\bf false}+{\bf
  true}$ for ${\bf x}$ and get the normal form $({\bf false}+{\bf
  true}) \otimes ({\bf false}+{\bf true})$, or start by using the fact
that all the functions defined in our language must be linear and get
$((\lambda {\bf x}\, ({\bf x} \otimes {\bf x})) \, {\bf false}) +
((\lambda {\bf x}\, ({\bf x} \otimes {\bf x})) \, {\bf true})$ and
finally the normal form $({\bf false} \otimes {\bf false}) + ({\bf
  true} \otimes {\bf true})$, leading to two different results. More
generally, faced with a term of the form $(\lambda {\bf x}\, {\bf t})
\, ({\bf u}+{\bf v})$, one could either start by substituting ${\bf
  u}+{\bf v}$ for ${\bf x}$, or start by applying the right-hand-side
linearity of the application, breaking the confluence of the calculus.
So that operations remain linear, it is clear that we must start by
developing over the $+$ first, until we reach a base vector and then
apply $\beta$-reduction. By base vector we mean a term which does not
reduce to a superposition. Therefore, we define a reduction strategy
where the $\beta$-reduction rule is restricted to cases where the
argument is a base vector, as formalized later.

With this restriction, we say that our language allows {\em copying}
but not {\em cloning} \cite{arrighidowek2,Altenkirch}.
It is clear that copying has all the expressiveness required in order to
express control flow, since it behaves exactly like the standard
$\beta$-reduction as long as the argument passed is not in a
superposition. This is the appropriate linear extension of the
$\beta$-reduction, philosophically it comprehends classical
computation as a (non-superposed) sub-case of linear-algebraic/quantum 
computation.

The same applies to erasing: the term $\lambda {\bf x} \lambda {\bf
y}~{\bf x}$ expresses the linear operator mapping the base vector
${\bf b}_i \otimes {\bf b}_j$ to ${\bf b}_i$. 
Again this is in contrast with other programming languages
where erasing is treated in a particular fashion whether for the
purpose of linearity of bound variables or the introduction of quantum
measurement.

\VML{\smallskip}\noindent
\subsection{Base (in)dependence}

The main conceptual difficulty when
seeking to let our calculus be higher-order is to understand how it
combines with this idea of {\em copying}, i.e., duplicating only base
vectors.  Terms of the form $(\lambda {\bf x}\,
({\bf x}\,{\bf x})) \, (\lambda {\bf x}\, {\bf v})$ raise
the important question of whether the $\lambda$-term $\lambda
{\bf x}\, {\bf v}$ must be considered to be a base vector or
not:
\VS{As we want 
higher-orderness in the traditional sense, i.e., 
$(\lambda{\bf x}\,{\bf t})\,(\lambda{\bf y}\,{\bf u})\lra{\bf
t}[\lambda{\bf y}\,{\bf u}/{\bf x}]$, abstractions must be the base
vectors.}

\VL{\noindent \begin{itemize}
\item we need to restrict
$(\lambda{\bf x}\,{\bf t})\,{\bf u}\lra{\bf t}[{\bf u}/{\bf x}]$ to ``base
vectors'';
\item
we want higher-order in the traditional sense
$(\lambda{\bf x}\,{\bf t})\,(\lambda{\bf y}\,{\bf u})\lra{\bf t}[\lambda{\bf y}\,{\bf u}/{\bf x}]$;
\item therefore abstractions must be the base vectors;
\item since variables will only ever be substituted by base vectors, they also are base vectors.
\end{itemize}
It is clear that there is a notion of privileged basis arising in the calculus,
but without us having to a priori choose a canonical basis (e.g. we do
not introduce some arbitrary orthonormal basis $\{\ket{i}\}$ all of a sudden --
i.e., we have nowhere specified a basis at the first-order level).}
The eventual algebraic consequences of this notion of a privileged basis arising only
because of the higher-order level are left as a topic for
further investigations.
An important intuition is that $(\lambda {\bf x}\,{\bf v})$ is not the
vector itself, but its classical description, i.e., the machine
constructing it -- hence it is acceptable to be able to copy $(\lambda
{\bf x}\,{\bf v})$ whereas we cannot clone ${\bf v}$. 
The calculus does exactly this distinction.

\VML{\smallskip}\noindent
\subsection{Infinities \& confluence}

It is possible, in our calculus, to define fixed point operators. For
instance for each term ${\bf b}$ we can define the term
$${\bf Y}_{\bf b} = 
\big((\lambda{\bf x}\,({\bf b}+({\bf x}\,{\bf x}))\big)
\big(\lambda{\bf x}\,({\bf b}+({\bf x}\,{\bf x})))\big)$$
Then the term 
${\bf Y}_{\bf b}$ reduces to ${\bf b}+{\bf Y}_{\bf b}$, i.e., the term reductions
generate a computable series of vectors $(n.{\bf b}+{\bf Y}_{\bf b})_n$ 
whose ``norm'' grows towards infinity. This was expected in the presence of both fixed points and linear algebra, 
but the appearance of such infinities entails the appearance of indefinite forms, which we must
handle with great caution. Marrying the full power of untyped $\lambda$-calculus, including fixed
point operators etc., with linear-algebra therefore jeopardizes the confluence
of the calculus, unless we introduce some further restrictions. 
\begin{exa}
If we took an unrestricted factorization rule 
$\alpha.{\bf t}+\beta.{\bf t} \lra (\alpha+\beta).{\bf t}$, then 
the term ${\bf Y}_{\bf b} - {\bf Y}_{\bf b}$ would reduce
to $(1 + (-1)) . {\bf Y}_{\bf b}$ and then ${\bf 0}$. It would also be reduce
to ${\bf b} + {\bf Y}_{\bf b} - {\bf Y}_{\bf b}$ and then to ${\bf b}$, breaking the
confluence.
\end{exa}
Thus, exactly like in elementary calculus $\infty - \infty$
cannot be simplified to $0$, we need to introduce a restriction
to the rule allowing to factor 
$\alpha.{\bf t}+\beta.{\bf t}$ into $(\alpha+\beta).{\bf t}$
to the cases where ${\bf t}$ is finite. But what do we mean by finite?
Notions of norm in the usual mathematical sense seem difficult to
import here. In order to avoid infinities we would like to ask that {\bf t} is
normalizable, but this is impossible to test in
general. Hence, we restrict further the rule 
$\alpha.{\bf t}+\beta.{\bf t} \lra (\alpha+\beta).{\bf t}$
to the case where the
term ${\bf t}$ is normal. It is quite striking to see how this restriction equates the algebraic notion of ``being normalized'' with the rewriting notion of ``being
normal''. 
The next three examples show that this indefinite form may pop up in some other, more hidden, ways.
\begin{exa}
Consider the term $(\lambda{\bf x}\,(({\bf x}\,\_) - ({\bf x}\,\_)))\,(\lambda
{\bf y}\, {\bf Y}_{\bf b})$ 
where $\_$ is any base vector, for instance ${\bf false}$.
If the term  $({\bf x}\,\_) - ({\bf x}\,\_)$ reduced to ${\bf 0}$ then
this term would both reduce to ${\bf 0}$ and also to 
${\bf Y}_{\bf b} - {\bf Y}_{\bf b}$, breaking confluence.
\end{exa}
Thus, the term ${\bf t}$ we wish to factor must also be closed, so that it does not contain any hidden infinity.
\begin{exa}
If we took an unrestricted rule 
$({\bf t} + {\bf u}) \, {\bf v}
\lra 
({\bf t} \, {\bf v}) + ({\bf u} \, {\bf v})$
the term $(\lambda{\bf x}\, ({\bf x}\,\_) -
\lambda{\bf x}\,({\bf x}\,\_))\,(\lambda {\bf y}\,{\bf Y}_{\bf b})$ would
reduce to ${\bf Y}_{\bf b} - {\bf Y}_{\bf b}$ and also to ${\bf 0}$, breaking
confluence.
\end{exa}
Thus we have to restrict the rule 
$({\bf t} + {\bf u}) \, {\bf v}
\lra 
({\bf t} \, {\bf v}) + ({\bf u} \, {\bf v})$
to the case where ${\bf t} +
{\bf u}$ is normal and closed.
\begin{exa}
If we took an unrestricted rule 
$(\alpha . {\bf u})\,{\bf v} \lra 
\alpha . ({\bf u}\,{\bf v})$ then 
the term $(\alpha.({\bf x}+{\bf y})) \, {\bf Y}_{\bf b}$ would reduce both to 
$(\alpha.{\bf x}+ \alpha.{\bf y}) \, {\bf Y}_{\bf b}$ and to 
$\alpha.(({\bf x}+{\bf y}) \, {\bf Y}_{\bf b})$, breaking confluence due to the previous restriction.
\end{exa}
Thus we have to restrict the rule 
$(\alpha . {\bf u})\,{\bf v} \lra 
\alpha . ({\bf u}\,{\bf v})$
to the case where ${\bf u}$ is
normal and closed.

This discussion motivates each of the restrictions $(*)-(****)$ in the
rules below. These restrictions are not just a fix: they are a way to
formalize vector spaces in the presence of limits/infinities. It may
come as a surprise, moreover, that we are able to tame these
infinities with this small added set of restrictions, and without any
need for context-sensitive conditions, as we shall prove in Section
\ref{confluence}.

\section{A higher-order language}\label{language}

We consider a first-order language, called {\em the language of
scalars}, containing at least constants $0$ and $1$ and binary
function symbols $+$ and $\times$. The {\em language of vectors} is
a two-sorted language, with a sort for vectors and a sort for
scalars, described by the following term grammar: 
$${\bf t}::= \quad {\bf x} \quad | \quad \lambda{\bf x}\, 
{\bf t}\quad |\quad {\bf t}~{\bf t}\quad |\quad{\bf 0}\quad |\quad
\alpha.{\bf t}\quad|\quad {\bf t}+{\bf t}\quad$$
where $\alpha$ is a term in the language of scalars.

In this paper we consider only semi-open terms, i.e., terms containing 
vector variables but no scalar variables. In particular all scalar
terms will be closed.

As usual we write ${\bf t}\,{\bf u}_1\, ...\, {\bf u}_n$ for 
$...({\bf t}\,{\bf u}_1)\, ...\, {\bf u}_n$. 

\begin{defi}[The system $L$]
\label{defsystemL}
Our small-step semantics is defined by the relation
$\lra_L$ where $L$ is the 
AC-rewrite system where the only AC-symbol is $+$ and the rules are 
those of $S$, a scalar rewrite system (see Definition \ref{scalarrs})
and the union of four groups of rules $E$, $F$, $A$ and $B$:\\
- Group $E$ -- elementary rules
\begin{align*}
{\bf u} + {\bf 0}  &\lra {\bf u},\\
0 . {\bf u} &\lra {\bf 0},\\
1 . {\bf u} &\lra {\bf u},\\
\alpha . {\bf 0} &\lra {\bf 0},\\
\alpha . (\beta . {\bf u}) &\lra (\alpha \times \beta). {\bf u},\\
\alpha . ({\bf u} + {\bf v}) &\lra \alpha . {\bf u} + \alpha . {\bf v}
\end{align*}
- Group $F$ -- factorisation
\begin{align*}
\alpha . {\bf u} + \beta . {\bf u} &\lra (\alpha + \beta). {\bf u},\qquad & (*)\\
\quad \alpha . {\bf u} + {\bf u} &\lra (\alpha + 1) . {\bf u}, & (*)\\
\quad {\bf u} + {\bf u} &\lra (1 + 1) . {\bf u}, & (*)
\end{align*}
- Group $A$ -- application\\
\begin{align*}
({\bf u} + {\bf v})~{{\bf w}} &\lra ({\bf u}~{\bf w})
+ ({\bf v}~{\bf w}), \qquad & (**)\\
{\bf w}~({\bf u} + {\bf v}) &\lra ({\bf w}~{\bf u}) + 
({\bf w}~{\bf v}), & (**)\\ 
(\alpha . {\bf u})~{\bf v} &\lra \alpha . ({\bf u}~{\bf v}), & (***)\\
\qquad {\bf v}~(\alpha . {\bf u}) &\lra \alpha . ({\bf v}~{\bf u}), & (***)\\
{\bf 0}~{\bf u} &\lra {\bf 0},\\
{\bf u}~{\bf 0} &\lra {\bf 0},
\end{align*}
- Group $B$ -- beta reduction\\
\begin{align*}
(\lambda {\bf x}~{\bf t})~{\bf b} \lra {\bf t}[{\bf b}/{\bf x}]\qquad \qquad (****) 
\end{align*}
And:\\
$(*)$ the three rules apply only if ${\bf u}$ is a closed $L$-normal
term.\\
$(**)$ the two rules apply only if ${\bf u} + {\bf v}$ is a closed $L$-normal
term.\\
$(***)$ the two rules apply only if ${\bf u}$ is a closed $L$-normal term.\\
$(****)$ the rule applies only when ${\bf b}$ is a ``base vector'' term, 
i.e., an abstraction or a variable.
\end{defi}
Notice that the restriction $(*)$, $(**)$ and $(***)$ are well-defined
as the terms to which the restrictions apply are smaller than the
left-hand side of the rule.\\ Notice also that the restrictions are
stable by substitution. Hence these conditional rules could be
replaced by an infinite number of non conditional rules, i.e., by
replacing the restricted variables by all the closed normal terms
verifying the conditions.\\ Finally notice how the rewrite system
$R=S\cup E\cup F\cup A$, taken without restrictions, is really just
the $V'$ (see Definition \ref{V'}) we have seen in Section
\ref{vecspaces}, i.e., an oriented version of the axioms of vector
spaces. Intuitively the restricted system defines a notion of vector
space with infinities. Rewrite rules with closedness conditions are not new
in the theory of $\lambda$-calculus (see, for instance, \cite{FernandezMackie}).

\VML{\smallskip}\noindent
\emph{Normal forms.} We have explained why abstractions ought to be considered as
``base vectors'' in our calculus. We have highlighted the presence of
non-terminating terms and infinities, which make it
impossible to interpret the calculus in your usual vector space
structure. The following \VML{two results show }\VS{result shows} that terminating closed terms on
the other hand can really be viewed as superpositions of abstractions.

\VML{\begin{prop}\label{abstractions}
An $L$-closed normal form, that is not a sum, a product by a
scalar, or the null vector, is an abstraction.
\end{prop}}

\VML{\proof{By induction over term structure. Let ${\bf t}$ be a closed
normal term that is not a sum, a product by a scalar,
or the null vector. The term ${\bf t}$ is not a variable because it is closed,
hence it is either an abstraction in which case we are done, or an
application. In this case it has the form
${\bf u}\,{\bf v}_1\,\ldots\,{\bf v}_n$ where ${\bf u},
{\bf v}_1, \ldots {\bf v}_n$ are normal and closed and $n$ is different
from $0$. Neither ${\bf u}$ nor ${\bf v}_1$ is a sum, a product by a
scalar, or the null vector since the term being normal we then could
apply rules of group $A$. Thus by induction hypothesis, both terms are
abstractions, thus the rule $B$ applies and the term ${\bf t}$ is not normal.\qed}}

\begin{prop}[Form of closed normal forms]\label{normalforms}
A $L$-closed normal form is either the null vector or of the form
$\sum_i \alpha_i. \lambda{\bf x}\,{\bf t}_i + \sum_i \lambda{\bf
x}\,{\bf u}_i $ where the scalars are different from $0$
and $1$ and the abstractions are all distinct.
\end{prop}

\VML{\proof{If the term is not the null vector it can be written as a sum
of terms that are neither ${\bf 0}$ nor sums. We partition these terms
in order to group those which are weighted by a scalar and those which
are not. Hence we
obtain a term of the form
$$\sum \alpha_i'.{\bf t'}_i + \sum {\bf u'}_i $$ where the terms
${\bf u'}_i$ are neither null, nor sums, nor weighted by a
scalar. Hence by Proposition \ref{abstractions} they are
abstractions. Because the whole term is normal the terms ${\bf t'}_i$
are themselves neither null, nor sums, nor weighted by a scalar since
we could apply rules of group $E$. Hence Proposition \ref{abstractions} also
applies.\qed}}

\section{Encoding classical and quantum computation}\label{encodings}

The restrictions we have placed upon our language are still more
permissive than those of the call-by-value $\lambda$-calculus, hence
any classical computation can be expressed in the linear-algebraic
$\lambda$-calculus just as it can in the call-by-value
$\lambda$-calculus. 

It then
suffices to express the three universal quantum gates ${\bf H},{\bf
Phase},{\bf Cnot}$ \cite{Boykin} which we will do next.

\VML{\smallskip} \noindent
\emph{Encoding booleans.}\, We encode the booleans as the first
and second projections, as usual in the classical $\lambda$-calculus:
$ {\bf true}\equiv\lambda {\bf x}\,\lambda {\bf y}\, {\bf x},\;{\bf false}\equiv\lambda {\bf x}\,\lambda {\bf y}\, {\bf y} $. Again, note that these are conceived as linear functions, the fact we
erase the second/first argument does not mean that the term should be
interpreted as a trace out or a measurement. \VML{Here is a standard example on how to use them:
\begin{align*}
{\bf Not}&\equiv \lambda {\bf y}\, \Big({\bf y}\,
{\bf false} \, {\bf true} \Big).
\end{align*}
Notice that this term expresses a unitary operator upon the vector space generated by ${\bf true}$ and ${\bf false}$, even if some subterms express non unitary ones.}

\VML{\smallskip} \noindent
\emph{Encoding unary quantum gates.}\, For the Phase gate the naive
encoding will not work, i.e.,
\begin{align*}
{\bf Phase}&\not\equiv \lambda {\bf y}\, \Big({\bf y}\,
(e^{i\frac{\pi}{4}}.{\bf true}) \, {\bf false} \Big)
\end{align*}
since by bilinearity this would give 
${\bf Phase}~{\bf false}\lra_L^* e^{i\frac{\pi}{4}}.{\bf false}$, 
whereas the Phase gate is supposed to place an $e^{i\frac{\pi}{4}}$ 
only on ${\bf true}$. The trick is to use
abstraction in order to retain the $e^{i\frac{\pi}{4}}$ phase on
${\bf true}$ only (where \_ is any base vector, for instance ${\bf false}$).
\begin{align*}
{\bf Phase}&\equiv \lambda {\bf y}\, \bigg(\Big({\bf y}\,\lambda
{\bf x}\,(e^{i\frac{\pi}{4}}.{\bf true})\,\lambda
{\bf x}\,{\bf false}\Big)\,\_\bigg)
\end{align*}
Now, the term ${\bf Phase}\,\,{\bf true}$ reduces in the following way
\VL{\begin{align*}
&\lambda {\bf y}\, \bigg(\Big({\bf y}\,\lambda
{\bf x}\,(e^{i\frac{\pi}{4}}.{\bf true})\,\lambda
{\bf x}\,{\bf false}\Big)\,\_\bigg)\,{\bf true}
\lra_L\Big({\bf true}\,\lambda
{\bf x}\,(e^{i\frac{\pi}{4}}.{\bf true})\,\lambda
{\bf x}\,{\bf false}\Big)\,\_ \\ 
&= \Big((\lambda {\bf x}\,\lambda
{\bf y}\, {\bf x})\,\lambda
{\bf x}\,(e^{i\frac{\pi}{4}}.{\bf true})\,\lambda
{\bf x}\,{\bf false}\Big)\,\_ 
\lra_L^*(\lambda {\bf x}\,(e^{i\frac{\pi}{4}}.{\bf true}))\,\_ 
\lra_L e^{i\frac{\pi}{4}}.{\bf true}
\end{align*}}
\VSM{$e^{i\frac{\pi}{4}}.{\bf true}$}
whereas the term ${\bf Phase}\,\,{\bf false}$ reduces as follows
\VL{\begin{align*}
&\lambda {\bf y}\, \bigg(\Big({\bf y}\,\lambda
{\bf x}\,(e^{i\frac{\pi}{4}}.{\bf true})\,\lambda
{\bf x}\,{\bf false}\Big)\,\_\bigg)\,{\bf false}
\lra_L \Big({\bf false}\,\lambda
{\bf x}\,(e^{i\frac{\pi}{4}}.{\bf true})\,\lambda
{\bf x}\,{\bf false}\Big)\,\_ \\ 
&= \Big((\lambda {\bf x}\,\lambda
{\bf y}\, {\bf y})\,\lambda
{\bf x}\,(e^{i\frac{\pi}{4}}.{\bf true})\,\lambda
{\bf x}\,{\bf false}\Big)\,\_ 
\lra_L^*(\lambda {\bf x}\,{\bf false})\,\_ 
\lra_L {\bf false}
\end{align*}}
\VSM{{\bf false}.}
This idea of using a dummy abstraction to restrict linearity can be
generalized and made more elegant with the following syntactic sugar: 
\begin{itemize}
\item ${\bf [t]}\equiv\lambda {\bf x}\, {\bf t}$. The effect of this {\em canon} $[.]$ is to associate to any state ${\bf t}$ a base vector ${\bf [t]}$. 
\item ${\bf \{t\}}\equiv{\bf t}\,\_$ where $\_$ is any closed normal base vector, for instance $\lambda {\bf x}\,{\bf x}$. The effect of this {\em uncanon} is to lift the canon, i.e., we have the derived rule ${\bf \{[t]\}}\lra_L {\bf t}$. 
\end{itemize}
Note that $\{.\}$ is a ``left-inverse'' of
$[.]$, but not a ''right inverse'', just like {eval} and $'$ (quote) in LISP. 
Again these hooks do not add anymore power to the
calculus, in particular they do not enable cloning. We cannot clone a
given state $\alpha.{\bf t}+\beta.{\bf u}$, but we can copy its
classical description $[\alpha.{\bf t}+\beta.{\bf u}]$. For instance the function $\lambda{\bf x} \, [{\bf x}]$ will never ``canonize'' anything else than a base vector, because of restriction $(****)$. 
The phase gate can then be written
\begin{align*}
{\bf Phase}&\equiv 
\lambda {\bf y}\, \big\{({\bf y}\,
[e^{i\frac{\pi}{4}}.{\bf true}])\,[{\bf false}]\big\} 
\end{align*}
For the Hadamard gate the game is just the same:
\begin{align*}
{\bf H}&\equiv 
\lambda {\bf y}\, \big\{{\bf y}\,
[\frac{\sqrt{2}}{2}.({\bf false}+{\bf true})]\,[\frac{\sqrt{2}}{2}.({\bf false}-{\bf true})]\big\}
\end{align*}

\VML{\smallskip}\noindent
\emph{Encoding tensors.} In quantum mechanics, vectors are put
together via the bilinear symbol $\otimes$. But because in our
calculus application is bilinear, the usual encoding of pairs does
just what is needed.
\begin{align*}
{\bf \otimes}&\equiv\lambda {\bf x}\,\lambda {\bf y}\, \lambda {\bf f}\, \big({\bf f}\,{\bf x}\,{\bf y}\big),\quad
{\bf \pi}_1\equiv\lambda {\bf p} ({\bf p}\,\lambda {\bf x}\,\lambda {\bf y}\, {\bf x}),\quad
{\bf \pi}_2\equiv\lambda {\bf p} ({\bf p}\,\lambda {\bf x}\,\lambda {\bf y}\, {\bf y}),\\
{\bf \bigotimes}&\equiv\lambda {\bf f}\,\lambda {\bf g}\, \lambda {\bf x}\,
\bigg(\otimes \, \big({\bf f}\,(\pi_1\,{\bf x})\big) \,
\big({\bf g}\,(\pi_2\,{\bf x})\big)\bigg)
\end{align*}
\noindent E.g. ${\bf H^{\otimes 2}}\equiv\big(\bigotimes {\bf H}\,{\bf H}\big)$. 
From there on the infix notation for tensors will be used, i.e.,
${\bf t}\otimes{\bf u}\equiv \otimes\,{\bf t}\,{\bf u},\quad{\bf t}\bigotimes{\bf u}\equiv\bigotimes\,{\bf t}\,{\bf u}.$

\VS{The Cnot gate can be defined in a similar way.}
\VML{\smallskip \noindent
\emph{Encoding the Cnot gate.} This binary gate is
essentially a classical gate, its encoding is standard. 
\begin{align*}
&{\bf Cnot}\equiv\lambda {\bf x}\,\Bigg( ({\bf \pi}_1 \,
{\bf x})\otimes \bigg( \Big( ({\bf \pi}_1 \, {\bf x}) \, \big({\bf Not}
\, ({\bf \pi}_2 \, {\bf x})\big)\Big) \, ({\bf \pi}_2 \, {\bf x}) \bigg)
\Bigg) 
\end{align*}}

\VML{\smallskip}\noindent
\emph{Expressing the Deutsch-Josza algorithm parametrically.} 

As discusses in Section \ref{mainfeatures}, an advantage of a
higher-order language is that it permits to express black-box
algorithms, such as the Deutsch-Josza algorithm, in a parametric
way. We show now how to encode this algorithm in Lineal.

Here is the well-known simple example of the Deutsch algorithm, which is 
the $n = 1$ case of the Deutsch-Josza algorithm
$${\bf Dj}_1\equiv\lambda {\bf x}\,\Bigg({\bf H^{\otimes
2}}\,\bigg({\bf x}\,\Big({\bf H^{\otimes 2}} \, \big({\bf false}\otimes{\bf true}\big)\Big)\bigg)
\Bigg)$$
But we can also express control structure and use them to express the
dependence of the Deutsch-Josza algorithm with respect to the size of
the input. Encoding the natural number $n$ as the Church numeral
${\bf n}\equiv\lambda {\bf x}\, \lambda {\bf f}\, ({\bf f}^{n}~{\bf x})$ the term
$({\bf n}~{\bf H}~\lambda {\bf y}\, ({\bf H}\bigotimes{\bf y}))$ reduces to
${\bf H}^{\otimes^{n+1}}$ and similarly the term 
$({\bf n}~{\bf true}~\lambda
{\bf y}\, ({\bf false}\otimes{\bf y}))$ 
reduces to 
${\bf false}^{\otimes^{n}} \otimes {\bf true}$. Thus the expression of the
Deutsch-Josza algorithm term of the introduction is now
straightforward\VSM{.}\VL{:
\begin{center}
{\small ${\bf Dj}\equiv\lambda {\bf n}\lambda {\bf x}\,\Bigg(({\bf n}~{\bf H}~\lambda {\bf y}\, ({\bf H}\bigotimes{\bf y}))\,\bigg({\bf x}\,\Big(({\bf n}~{\bf H}~\lambda {\bf y}\, ({\bf H}\bigotimes{\bf y})) \, \big({\bf n}~{\bf true}~\lambda
{\bf y}\, ({\bf false}\otimes{\bf y})\big)\Big)\bigg)
\Bigg).$}
\end{center}
}

\VML{\smallskip \noindent
\emph{Infinite dimensional operators.}
Notice that our language enables us to express operators independently
of the dimension of the space they apply to, even when this
dimension is infinite. For instance the identity operator is not
expressed as a sum of projections whose number would depend on the
dimension of the space, but as the mere lambda term $\lambda {\bf x}\,{\bf
x}$. In this sense our language is a language of infinite dimensional
computable linear operators, in the same way that matrices are a
language of computable finite dimensional linear operators.}

\section{Confluence}\label{confluence}

The main theorem of this paper is the confluence of the system $L$. Along the way, we will also prove the confluence of the unrestricted systems $V$ and $V'$ which we introduced for the sake of our computational definition of vector spaces.
\VML{This section is quite technical. A reader who is not familiar with
rewriting techniques may be happy with reading just Definition \ref{defconfluence}
and Theorem \ref{thconfluence}, and then skipping to Section \ref{currentworks}.
A reader with an interest in such techniques may on the other hand find 
the architecture of the proof quite useful.} 
\VL{The rationale is as follows:
\begin{itemize}
\item There is only a very limited set of
techniques available for proving the confluence of non-terminating
rewrite systems (mainly the notions of parallel reductions and strong
confluence). Hence we must distinguish the non-terminating rules,
which generate the infinities (the $B$ rule), from
the others (the $R$ rules) and show that they are
confluent on their own;
\item We then must
show that the terminating rules and the non-terminating rules commute,
so that their union is confluent. (The conditions on
$B,F,A$ are key to obtaining the commutation. Without them, both
subsets are confluent, but not their union.)

\item The rules of $R$ being terminating, the
critical pairs lemma applies. The critical pairs can be checked
automatically for the non-conditional rules, 
but the conditional ones must be checked by hand.
Hence we must distinguish the non-conditional rules
($E$ rules), from the others (the $F$ and $A$ rules).

\item In order to handle the parametricity with respect to 
the scalars rewrite system, we shall introduce a 
new technique called the avatar's lemma.
\end{itemize}}

\noindent The first step is to prove the confluence of the
system $R=S \cup E \cup F \cup A$, i.e., the system $L$ 
of Definition \ref{defsystemL}, minus the
rule $B$. 
To prove the confluence of this system 
we shall prove that of the system $R_0 = S_0 \cup E \cup F \cup A$ where $S_0$ 
is a small avatar of $S$, namely the simplest possible scalar rewrite system. 
Then we use the avatar lemma 
to extend the result 
from $S_0$ to $S$, hereby obtaining the confluence of $R = S \cup E \cup F \cup A$.

\begin{defi} [The rewrite system $S_0$]
The system $S_0$ is formed by the rules
$$0 + \alpha \lra \alpha$$
$$0 \times \alpha \lra 0$$
$$1 \times \alpha \lra \alpha$$
$$\alpha \times (\beta + \gamma) \lra (\alpha \times \beta) + (\alpha \times \gamma)$$
where $+$ and $\times$ are AC-symbols.
\end{defi}

To be able to use a critical pair lemma in an AC context, 
we shall use a well-known technique, detailed in the Section \ref{extension},
and introduce addenda $S_{0ext}$, $E_{ext}$ and $F_{ext}$ to the systems
$S_0$, $E$ and $F$, with some extra 
rules called {\em extension} rules, but with a more restricted form of 
AC-rewriting and the system $R_{0ext} = S_{0ext} \cup E_{ext} \cup F_{ext} \cup A$.

The second step towards our main goal is
show that the
$B^\parallel$ rule, the parallel version of the rule $B$ defined in Definition
\ref{defBpar}, 
is strongly confluent on the term algebra, and
commutes with $R^*$, hence giving the confluence of $L$ (Section
\ref{betaconfluence}).

As the system $R=S \cup E \cup F \cup A$ does
not deal at all with abstractions and bound variables, we have,
throughout this first part of the proof, considered $\lambda {\bf x}$
as a unary function symbol and the bound occurrences of ${\bf x}$ as
constants. This way we can safely apply known theorems about
first-order rewriting.

\subsection{Reminder on rule extensions and the critical pairs lemma}
\label{extension}

The term $(({\bf a}+{\bf 2.a})+{\bf b})+{\bf c}$ reduces to the term
$((2+1).{\bf a} + {\bf b})+{\bf c}$, as it contains a subterm ${\bf
a}+{\bf 2.a}$ that is AC-equivalent to an instance of the left
hand side of the rule $\alpha . {\bf u} + {\bf u} \lra (\alpha + 1) . {\bf
u}$.  The term $((2 . {\bf a}+{\bf b})+{\bf a})+{\bf c}$ does not
contain such a subterm, yet it can be reduced, because
it is itself AC-equivalent to the term $((2. {\bf a}+{\bf a})+{\bf
  b})+{\bf c}$ that contain a subterm that is an instance of the left
hand side of this rule.  This suggests that there is a more local way
to define reduction modulo AC where the first reduction is possible
but not the second.

Unfortunately, the critical pair lemma for AC-rewrite system gives
the confluence of this local version of AC-reduction system and not 
the global one we
are interested in.
This problem has been solved by \cite{PetersonStickel,JouannaudKirchner} 
that show that the globally AC reduction relation is confluent if 
the locally AC reduction relation of an extended rewrite system is confluent. 

\couic{
\begin{defi}[The locally AC-reduction relation $\lra_{X, loc}$]
Let $X$ be a rewrite system, we define the relation $\lra_{X, loc}$ as
follows $t \lra_{X, loc} u$ if there exists an occurrence $p$ in ${t}$, a
rewrite rule $l \lra r$ in $X$ and a substitution $\sigma$ such that
$t_{|p} =_{AC} \sigma l$ and $u =_{AC} t[p \leftarrow \sigma r]$.
\end{defi}
}

\begin{defi}[The extension rules]
Let $X$ be a AC-rewrite system with AC symbols $f_1,\ldots,f_n$. We define the AC-rewrite system $X_{ext}$ as containing the same AC symbols as $X$, the same rules as $X$, plus the rules $f_i(t,x)\lra f_i(u,x)$ for each rule $t\lra u$ of $X$ where the head symbol of $t$ is $f_i$.
\end{defi}

\begin{prop}[$R_{0ext}$]
The system $S_{0ext}$ is formed by the rules of $S_0$ and the rule
$$(0 + \alpha) + \chi \lra \alpha + \chi$$
$$(0 \times \alpha) \times \chi \lra 0 \times \chi$$
$$(1 \times \alpha) \times \chi \lra \alpha \times \chi$$
$$(\alpha \times (\beta + \gamma)) \times \chi 
\lra ((\alpha \times \beta) + (\alpha \times \gamma)) \times \chi$$
The system $E_{ext}$ is formed by the rules of $E$ and the rule:
$$({\bf u}+{\bf 0})+{\bf x}\lra {\bf u}+{\bf x}$$
The system $F_{ext}$ is formed by the rules of $F$ and these
three rules:
$$(\alpha . {\bf u} + \beta . {\bf u}) + {\bf x} \lra (\alpha +
\beta) . {\bf u} + {\bf x}\quad (*)$$
$$(\alpha . {\bf u} + {\bf u}) + {\bf x} \lra (\alpha + 1) . {\bf u}
+ {\bf x}\quad (*)$$
$$({\bf u} + {\bf u}) + {\bf x} \lra (1 + 1) . {\bf u} +
{\bf x}\quad (*)$$
where $(*)$ imposes the three rules apply only if ${\bf u}$ is a closed 
normal term.
The system $A_{ext}$ is $A$.\\
The system $R_{0ext}$ is $S_{0ext} \cup E_{ext} \cup F_{ext} \cup A$. 
\end{prop}
As usual we write
${\bf t}\lra_X^*{\bf u}$ if and only if ${\bf t}={\bf u}$ or
${\bf t}\lra_X \ldots \lra_X {\bf u}$. We also write 
${\bf t}\lra_X^?{\bf u}$ if and only if ${\bf t} = {\bf u}$ or
${\bf t}\lra_X{\bf u}$
${\bf t} \lra_{X;Y} {\bf u}$
if there exist a ${\bf v}$ such that 
${\bf t} \lra_{X} {\bf v} \lra_{Y} {\bf u}$, 
and 
${\bf t} \lra_{X\downarrow} {\bf u}$
if ${\bf t} \lra_{X}^* {\bf u}$ and ${\bf u}$ is normal for the relation $X$.

\subsection{Termination}

\begin{prop}
The systems $S_{0ext}$ terminates.
\end{prop}
\proof{Consider the following interpretation (compatible with AC)
$$|x|_s = |0|_s = |1|_s = 2$$
$$|t + u|_s = 1 + |t|_s + |u|_s$$
$$|t \times u|_s = |t|_s \times |u|_s$$
Each time a term $t$ rewrites to a term $t'$ we have
$|t|_s > |t'|_s$. Hence, the system terminates.\qed}

\begin{prop}\label{prop:terminates}
The systems $R_{0ext}$, $R_{ext}$, $S\cup V$ and $S\cup V'$ terminate.
\end{prop}
\VL{\proof{[\emph{The system $E_{ext} \cup F_{ext} \cup A$ terminates}]\\
Consider the following interpretation (compatible with AC)
$$|({\bf u} ~ {\bf v})| = (3 |{\bf u}| + 2)(3 |{\bf v}| + 2)$$
$$|{\bf u} + {\bf v}| = 2 + |{\bf u}| + |{\bf v}|$$
$$|\alpha . {\bf u}| = 1 + 2 |{\bf u}|$$
$$|{\bf 0}| = 0$$
Each time a term ${\bf t}$ rewrites to a term ${\bf t'}$ we have
$|{\bf t}| > |{\bf t'}|$. Hence, the system terminates.

\VML{\smallskip}\noindent 
[\emph{The system $R_{0ext}$ terminates}]\\
The system $R_{0ext}$ is $S_{0ext} \cup E_{ext} \cup F_{ext} \cup A$. 
It is formed of two
subsystems $S_{0ext}$ and $E_{ext} \cup F_{ext} \cup A$. 
By definition of the function $|~|$, if a term ${\bf t}$ $S_{0ext}$-reduces 
to a term
${\bf t'}$ then $|{\bf t}| = |{\bf t'}|$.
Consider a $R$-reduction sequence. At each $E_{ext} \cup F_{ext} \cup A$-reduction
step, the measure of the term strictly decreases and at each
$S_{0ext}$-reduction step it remains the same. Thus there are only a finite
number of $E_{ext} \cup F_{ext} \cup A$-reduction steps in the sequence and, as
$S_{0ext}$ terminates, the sequence is finite.

The same argument applies for $R_{ext}$, $S\cup V$ and $S\cup V'$, with respect to $S$ instead of $S_{0ext}$.
}}
\VM{\proof{See the long version of the paper \cite{arrighidowek3}.\qed}}

\subsection{Critical pairs}

\VML{
\begin{defi}[Confluence and local confluence]\label{defconfluence}
A relation $X$ is said to be confluent if whenever $t\lra_X^*
u$ and $t\lra_X^* v$, there exists a term $w$ such that $u \lra_X^* w$
and $v \lra_X^* w$. A relation $X$ is said to be locally confluent if whenever $t\lra_X u$ and $t\lra_X v$, there exists a term $w$ such that $u \lra_X^* w$ and $v \lra_X^* w$. 
\end{defi}

\begin{defi}[Critical pair]
Let $l \lra r$ and $l' \lra r'$ be two rewrite rules of an AC-rewrite
system $X$, let $p$ be an occurrence in $l$ such that $l_{|p}$ is not
a free variable. Let $\sigma$ be an AC-unifier for $l_{|p}$ and $l'$,
the pair $\big(\sigma r, \sigma (l[p \leftarrow r'])\big)$ is a {\em
critical pair} of the the rewrite system $X$.
\end{defi}

\begin{prop}[Peterson-Stickel Theorem]\label{cplemma}
If $\lra_{X_{ext}}$ terminates and for each critical pair $({t},{u})$ of $X_{ext}$ there exists a term ${w}$ such that ${t}\lra_{X_{ext}}^* {w}$ 
${u}\lra_{X_{ext}}^* {w}$.
Then the relation $\lra_{X}$ is confluent.
\end{prop}
\proof{See \cite{PetersonStickel}, Theorems 10.5., 9.3 and 8.9.\qed}

\VS{The notions of confluence, strong confluence, local confluence and critical pair are
as usual. We use the critical pair lemma and the following lemma that
is a consequence of the Theorems 8.9, 9.3 and 10.5 of
\cite{PetersonStickel}.}

\begin{prop}
\label{N0localconfluent}
The system $S_0 \cup E$ is confluent.\\
The systems $S_0\cup V$ and $S_0\cup V'$ are confluent. 
\end{prop}
\proof{
First notice that $(S_0 \cup E)_{ext}=(S_{0ext} \cup E_{ext})$ and that this system terminates (see Proposition \ref{prop:terminates}). Thus by proposition \ref{cplemma} all we need to do is to check that all the critical pair close.
As these rules are not conditional, we have used the system CIME \cite{cime}
to check this automatically. The same applies to $S_0\cup V_{ext}$ and $S_0\cup {V'}_{ext}$.\qed 
}

\begin{prop}
\label{fact}
The system $S_0 \cup E \cup F$ is confluent.
\end{prop}
\VL{\proof{ First notice that the system $(S_{0ext} \cup E\cup F)_{ext} = S_{0ext}
\cup E_{ext}\cup F_{ext}$ and that this system terminates (see
Proposition \ref{prop:terminates}). Thus by proposition
\ref{cplemma} all we need to do is to check that all the critical
pair close.  If both rules used are rules of the system $S_{0ext} \cup
E_{ext}$, then the critical pair closes by Proposition
\ref{N0localconfluent}.  There are no critical pairs between $S_{0ext}$
and $E_{ext}\cup F_{ext}$.  Thus, all we need to check are the
critical pairs between one rule of $E_{ext}$ and one of $F_{ext}$
or one rule of $F_{ext}$ and one of $F_{ext}$.

To find the critical pairs, we used CIME \cite{cime}. There are
251 critical pairs, not taking the fact that some rules are conditional and thus may not apply. 
Indeed, among these critical pairs 81 do not verify the conditions of the rules
$F_{ext}$. For instance, the pair given by CIME
$$\alpha . ({\bf 0} + {\bf u}) +  {\bf u} 
\lla 
{\bf 0} + \alpha . ({\bf 0} + {\bf u}) +  {\bf u} 
\lra 
(1 + \alpha). ({\bf 0} + {\bf u}) $$
does not verify the condition because ${\bf u} + {\bf 0}$ is
never closed normal: we do not need to close this pair because our conditions forbid that it opens.

We need to check the 170 other pairs by hands. Some close easily, for instance 
the pair
$${\bf 0} + \alpha . {\bf u} 
\lla 
0 . {\bf u} + \alpha . {\bf u} 
\lra 
(0 + \alpha) . {\bf u}$$
closes on $\alpha . {\bf u}$.

Some other require a short analysis of the conditions. For instance, for the pair
$$\alpha . {\bf u} + \alpha . {\bf v} + \beta . ({\bf u} + {\bf v}) 
\lla 
\alpha . ({\bf u} + {\bf v}) + \beta . ({\bf u} + {\bf v}) 
\lra 
(\alpha + \beta) . ({\bf u} + {\bf v})$$
the fact that we have been able to factor 
$\alpha . ({\bf u} + {\bf v}) + \beta . ({\bf u} + {\bf v})$
into
$(\alpha + \beta) . ({\bf u} + {\bf v})$
shows that the term
${\bf u} + {\bf v}$ is closed normal, 
thus the terms ${\bf u}$ and ${\bf v}$ are closed normal, 
which permits to reduce both terms to 
$(\alpha + \beta) . {\bf u} + (\alpha + \beta) . {\bf v}$.
All the cases are analyzed in \cite{251}.\qed}}

\VSM{
\proof{This system is made of two subsystems : $S \cup E$ and $F_{ext}$. 
To prove that it is locally confluent, we prove that all critical
pairs close. 
We used an AC-unification algorithm to compute these critical pairs.
If both rules used are rules of the system $S \cup 
E$, then the critical pair closes by Proposition \ref{Nlocalconfluent}. 
We check the 251 other critical pairs by hand. The detail can be found
in 
the long version of the paper \cite{arrighidowek3}.}}

\begin{prop}
\label{bubu}
The system $R_0 = S_0 \cup E \cup F \cup A$ is confluent.
\end{prop}
\VL{\proof{
First notice that $R_{0ext}=S_{0ext} \cup E_{ext} \cup F_{ext}\cup A$ and that this system terminates (see Proposition \ref{prop:terminates}). Thus by proposition \ref{cplemma} all we need to do is to check that all the critical pair close.
If both rules used are rules of the system $S_{0ext} \cup 
E_{ext} \cup F_{ext}$, then the critical pair closes by Proposition
\ref{fact}. 
It is not possible that the top-level rule is in $S_{0ext} \cup E_{ext} \cup 
F_{ext}$ and the other in $A$ since the rules of 
$S_{0ext} \cup E_{ext} \cup F_{ext}$ 
do not contain any application. 
Thus the top-level rule must be in $A$ and 
the 
$(S_{0ext} \cup E_{ext} \cup F_{ext})$-reduction must be performed in a non-toplevel
non-variable subterm of the left-hand-side of a rule of $A$. By inspection of the left-hand-sides of rules
$S_{0ext} \cup E_{ext} \cup F_{ext}$ the
subterm must be of the form ${\bf u} + {\bf v}$, $\alpha . {\bf u}$ or
${\bf 0}$. But this subterm cannot be of the form ${\bf u} + {\bf v}$, because, 
by restriction (**), the term itself would not be $A$-reducible. 
It cannot be ${\bf 0}$ since this term is normal. Thus it is of the form 
$\alpha . {\bf u}$. As there are five rules reducing a term of this
form, there are 10 critical pairs to check.  
Because of the conditionality of the rewrite system we check them by hand.
\begin{description}
\item[Pair 1] 
$
{\bf 0} \, {\bf v} 
\lla
(0.{\bf u}) \, {\bf v} 
\lra 
0.({\bf u} \, {\bf v})
$: this critical pair closes on ${\bf 0}$.

\item[Pair 2] 
$
{\bf u} \, {\bf v} 
\lla
(1.{\bf u}) \, {\bf v}
\lra 
1.({\bf u} \, {\bf v})
$: 
this critical pair closes on ${\bf u}\,{\bf v}$.

\item[Pair 3] 
$
{\bf 0} \, {\bf v} 
\lla 
(\alpha.{\bf 0}) \, {\bf v}
\lra \alpha.({\bf 0} \, {\bf v})
$:
this critical pair closes on ${\bf 0}$.

\item[Pair 4] 
$
((\alpha\times \beta).{\bf u}) \, {\bf v} 
\lla
(\alpha.(\beta.{\bf u})) \, {\bf v}
\lra 
\alpha.((\beta.{\bf u}) \, {\bf v})$:
the term ${\bf u}$ is closed and normal by $(***)$. 
Hence, the critical pair closes on 
$(\alpha \times\beta).({\bf u}\,{\bf v})$.

\item[Pair 5] 
$
(\alpha.{\bf u} + \alpha.{\bf v}) \, {\bf w} 
\lla
(\alpha.({\bf u} + {\bf v})) \, {\bf w}
\lra 
\alpha.(({\bf u} + {\bf v}) \, {\bf w})$:
the term ${\bf u}+{\bf v}$ is closed and normal. Hence, by
Proposition \ref{normalforms} it is of the form $\sum_i
\beta_i.{\bf a}_i + \sum_i {\bf b}_i$. Therefore the top reduct
reduces to $(\sum_i (\alpha\times \beta_i)\downarrow .{\bf a}_i +
\sum_i \alpha.{\bf b}_i) \, {\bf w}$, where $\downarrow$ denotes
normalization by $S_{0ext}$. We treat only the case where the terms $(l\times
\beta_i)\downarrow$ are neither $0$ nor $1$, the other cases being
similar. Hence, we can apply rules of group $A$ yielding $\sum_i
(\alpha\times \beta_i)\downarrow .({\bf a}_i \, {\bf w})+ \sum_i
\alpha.({\bf b}_i\, {\bf w})$. It is routine to check that the
bottom reduct also reduces to this term.

The five next critical pairs are the symmetrical cases, permuting the left and right-hand-sides of the application.
\end{description}

\noindent
Now, when both rules are in the group $A$, there are 9 critical
pairs to check.
\begin{description}
\item[Pair 11] 
$ 
{\bf u} ({\bf w} + {\bf x}) + {\bf v} ({\bf w} + {\bf x}) 
\lla
({\bf u} + {\bf v}) ({\bf w} +
{\bf x}) \lra ({\bf u} + {\bf v}) {\bf w} + ({\bf
u} + {\bf v}) {\bf x}$:
as ${\bf u} + {\bf v}$ and ${\bf w} + {\bf x}$ are
normal and closed, so are ${\bf u}$, ${\bf v}$, ${\bf w}$
and ${\bf x}$.  Hence the critical pair closes on ${\bf u}
{\bf w} + {\bf u} {\bf x} + {\bf v} {\bf w} +
{\bf v} {\bf x}$.

\item[Pair 12]
$
{\bf u} (\alpha.{\bf w}) + {\bf v} (\alpha.{\bf w}) 
\lla
({\bf u} + {\bf v}) (\alpha.{\bf w}) \lra \alpha. (({\bf u} + {\bf v}) . {\bf w})
$:
as, by $(**)$ and $(***)$, ${\bf u}
+ {\bf v}$ and ${\bf w}$ are closed normal terms , so are
${\bf u}$ and ${\bf v}$.  Thus the top reduct further reduces to
$\alpha.({\bf u} \, {\bf w}) + \alpha . ({\bf v} \,
{\bf w})$ and the bottom reduct further reduces to $\alpha. (({\bf
u} \, {\bf w}) + ({\bf v} \, {\bf w}))$ and both terms
reduce to $\alpha.({\bf u} \, {\bf w}) + \alpha.({\bf v}
\, {\bf w})$.

\item[Pair 13]
$
{\bf 0} 
\lla
({\bf u} + {\bf v}) {\bf 0} 
\lra 
({\bf u} {\bf 0}) + ({\bf v} {\bf 0})$:
this critical pair closes on
${\bf 0}$.

\item[Pair 14]
$
\alpha.({\bf u} ({\bf v} + {\bf w})) 
\lla
(\alpha. {\bf u})({\bf v} + {\bf w}) 
\lra 
(\alpha. {\bf u}) {\bf v} + (\alpha. {\bf u}) {\bf w}$:
the terms ${\bf u}$ and ${\bf v} + {\bf w}$ are closed normal.  
Thus, the
top reduct further reduces to $\alpha .({\bf u} {\bf v} + {\bf u}
{\bf w})$ and the bottom reduct to $\alpha. ({\bf u} {\bf v}) +
\alpha. ({\bf u} {\bf w})$. Hence the critical pair closes on $\alpha. ({\bf u}
{\bf v}) + \alpha. ({\bf u} {\bf w})$.

\item[Pair 15]
$\alpha.({\bf u} (\beta. {\bf v})) 
\lla
(\alpha. {\bf u}) (\beta . {\bf v}) 
\lra 
\beta. ((\alpha. {\bf u}) {\bf v})$:
as ${\bf u}$ and ${\bf v}$ are closed normal, the first
term reduces to $\alpha.(\beta. ({\bf u} {\bf v}))$ and the
second to $\beta. (\alpha.({\bf u} {\bf v}))$ and both terms
reduce to $(\alpha \times\beta) . ({\bf u} {\bf v})$.

\item[Pair 16] 
$\alpha.({\bf u} {\bf 0}) 
\lla
(\alpha. {\bf u}) {\bf 0}\lra {\bf 0}$:
this critical pair closes on
${\bf 0}$.

\item[Pair 17]
$
{\bf 0} \lla
{\bf 0} ({\bf u} + {\bf v}) \lra
({\bf 0} {\bf u}) + ({\bf 0} {\bf v})$:
this critical pair closes on
${\bf 0}$.

\item[Pair 18]
${\bf 0} 
\lla
{\bf 0} (\alpha.{\bf u}) \lra \alpha.({\bf 0}
{\bf u})$:
this critical pair closes on
${\bf 0}$.

\item[Pair 19] 
${\bf 0} 
\lla
{\bf 0} {\bf 0} \lra {\bf 0}$:
this critical pair closes on
${\bf 0}$.\qed
\end{description}
}}

\VM{
\proof{
This system is made of two subsystems: $S \cup E \cup F_{ext}$ and $A$.  
To prove that it is locally confluent, we prove
that all critical pairs close. If both rules used are rules of the system $S \cup 
E \cup F_{ext}$, then the critical pair closes by Proposition
\ref{fact}. 
It is not possible that the top-level rule is in $S \cup E \cup 
F_{ext}$ and the other in $A$ since the rules of 
$S \cup E \cup F_{ext}$ 
do not contain any application. 
Thus the top-level rule must be in $A$ and 
the 
$(S \cup E \cup F_{ext})$-reduction must be performed in a non-toplevel
non-variable subterm of the left-hand-side of a rule of $A$. By inspection of the left-hand-sides of rules
$S \cup E \cup F_{ext}$ the
subterm must be of the form ${\bf u} + {\bf v}$, $\alpha . {\bf u}$ or
${\bf 0}$. But this subterm cannot be of the form ${\bf u} + {\bf v}$, because, 
by restriction (**), the term itself would not be $A$-reducible. 
It cannot be ${\bf 0}$ since this is normal. Thus it is of the form 
$\alpha . {\bf u}$. As there are five rules reducing a term of this
form, there are 10 critical pairs to check. When both rules are in the
group $A$, there are nine critical pairs to check. The detail can be
found in 
the long version of the paper
\cite{arrighidowek3}.}}
\VS{\proof{Similar to above. See \cite{arrighidowek3}.\qed}}

}

\subsection{The avatar lemma}\label{avatar} \label{appcp1}

\begin{defi}[Subsumption]
A terminating and confluent relation $S$ {\em subsumes} a relation
$S_0$ if whenever $t\lra_{S_0} u$, $t$ and $u$ have the same $S$-normal
form.
\end{defi}

\begin{defi}[Commuting relations]
Two relations $X$ and $Y$ are said to be commuting if whenever $
t\lra_X u$ and $t\lra_Y v$, there exists a term $w$ such that $u
\lra_Y w$ and $v \lra_X w$.
\end{defi}

\begin{prop}[The avatar lemma]\cite{arrighidowek2}
\label{Lemma}
Let $X$, $S$ and $S_0$ be three relations defined on a set such that:
\begin{itemize}
\item $S$ is terminating and confluent;
\item $S$ subsumes $S_0$;
\item $S_0 \cup X$ is locally confluent;
\item $X$ commutes with $S^{*}$.
\end{itemize}
Then, the relation $S \cup X$ is locally confluent.
\end{prop}
\VL{\proof{
\noindent[\emph{$X$ can be simulated by $X;S\downarrow$}].\\
If $t \lra_X u$ and $t \lra_{S^\downarrow} v$, 
then there exists $w$ such that $u \lra_{S^\downarrow} w$
and $v \lra_{X;S^\downarrow} w$. 
Indeed by commutation of $X$ and $S^*$ there exists $a$ such that 
$u \lra_{S^*} a$ and $v \lra_{X} a$. 
Normalizing $a$ under $S$ yields the $w$.

\VML{\smallskip}\noindent 
[\emph{$S_0 \cup X$ can be simulated by $(X;S\downarrow)^{?}$}].

If $t \lra_{S_0 \cup X} u$ and $t \lra_{S^\downarrow} v$, 
then there exists $w$ such that $u \lra_{S^\downarrow} w$
and $v \lra_{X;S^\downarrow}^{?} w$. 
Indeed if $t\lra_{S_0} u$ this is just subsumption, else the first point of this proof applies.

\VML{\smallskip}\noindent 
[\emph{$S \cup X$ can be simulated by $(X;S\downarrow)^{?}$}].

If $t \lra_{S \cup X} u$ and $t \lra_{S^\downarrow} v$, 
then there exists $w$ such that $u \lra_{S^\downarrow} w$
and $v \lra_{X;S^\downarrow}^{?} w$. 
Indeed if $t \lra_{S} u$ this is just the normalization of $S$, else the first point of this proof applies.

\VML{\smallskip}\noindent 
[\emph{$X;S\downarrow$ is locally confluent}].

If $t \lra_{X;S^\downarrow} u$ and $t \lra_{X;S^\downarrow} v$, 
then there exists $w$ such that $u \lra_{X;S^\downarrow}^* w$
and $v \lra_{X;S^\downarrow}^* w$. 
Indeed if $t \lra_{X} a \lra_{S^\downarrow} u$ 
and $t \lra_{X} b \lra_{S^\downarrow} v$ we know from the local confluence of 
$S_0 \cup X$ that there exists $c$
such that $a \lra_{S_0 \cup X}^* c$ and $b \lra_{S_0 \cup X}^* c$.
Normalizing $c$ under $S$ yields the $w$. 
This is because by the second point of the proof $u \lra_{X;S^\downarrow}^* w$
and $v \lra_{X;S^\downarrow}^* w$.

\VML{\smallskip}\noindent 
[\emph{$S \cup X$ is locally confluent}].

If $t \lra_{S \cup X} u$ and $t \lra_{S \cup X} v$, then
there exists $w$ such that $u \lra_{S \cup X}^* w$ and
$v \lra_{S \cup X}^* w$. Indeed call $t^\downarrow$,
$u^\downarrow$, $v^\downarrow$ the $S$ normalized version of
$t$, $u$, $v$. By the third point of our proof we have
$t^\downarrow \lra_{X;S^\downarrow}^{?} u^\downarrow$ and
$t^\downarrow \lra_{X;S^\downarrow}^{?} v^\downarrow$. By the
fourth point of our proof there exists $w$ such that
$u^\downarrow \lra_{X;S^\downarrow}^* w$ and
$v^\downarrow \lra_{X;S^\downarrow}^* w$.\qed}}

\begin{prop}
\label{Nlocalconfluent}\label{Rconf}
For any scalar rewrite system $S$ the systems $R=S \cup E \cup F \cup A$, 
$S\cup V$ and $S\cup V'$ are confluent.
\end{prop}
\VML{\proof{The system $S$ is confluent and terminating because it is a
scalar rewrite system. The system $S$ subsumes $S_0$ because $S$ is a
scalar rewrite system.  From Proposition \ref{bubu}, the
system $R_0 = S_0 \cup E \cup F \cup A$ is confluent.  Finally, the
system $E \cup F \cup A$ commutes with $S^*$. Indeed, each rule of 
$E \cup F \cup A$ commutes with $S^*$ 
as each subterm of sort scalar in the left member of a rule
is either a variable or $0$ or $1$, which are normal
forms.  We conclude with Proposition \ref{Lemma} that 
$R=S\cup E \cup F \cup A$ is locally confluent. Hence as it is terminating it is confluent by Newman's lemma \cite{Newman}.The same argument applies for $S\cup V$ and $S\cup V'$.\qed}}

\subsection{The system $L$} \label{betaconfluence}

We now want to prove that the system $L$ is
confluent. With the introduction of the rule $B$, we lose
termination, hence we cannot use Newman's lemma \cite{Newman} anymore. 
Thus we shall use for this last part techniques coming from the proof of 
confluence of the $\lambda$-calculus and prove that the parallel version 
of the $B$ rule is strongly confluent. In our case as we have to
mix the rule $B$ with $R$ we shall also prove that it 
commutes with $\lra_R^{*}$.

\VML{\begin{defi}[Strong confluence]
A relation $X$ is said to be strongly confluent if whenever $t \lra_X
u$ and $t \lra_X v$, there exists a term $w$ such that $u \lra_X w$
and $v \lra_X w$.
\end{defi}}

\begin{defi}[The relation $\lra_{B}^{\parallel}$]
\label{defBpar}
The relation $\lra_{B}^{\parallel}$ is the smallest reflexive congruence such
that if ${\bf u}$ is a base vector, ${\bf t}~\lra_{B}^{\parallel}~{\bf t'}$
and ${\bf u}~\lra_{B}^{\parallel}~{\bf u'}$ then  
$$(\lambda {\bf x}~{\bf t})~{\bf u}~\lra_{B}^{\parallel}~{\bf t'}[{\bf u'}/{\bf x}]$$
\end{defi}

\VML{
\begin{prop}
\label{substitution}
If ${\bf v}_1 \lra_R^{*} {\bf w}_1$ 
then 
${\bf v}_1[{\bf b}/{\bf x}]
 \lra_R^{*} {\bf w}_1[{\bf b}/{\bf x}]$, where ${\bf b}$ is a base vector.
\end{prop}
\VL{\proof{If the reduction of ${\bf v}_1$
to ${\bf w}_1$ involves an application of a conditional rule, 
then the condition is preserved on 
${\bf v}_1[{\bf v}_2/{\bf x}]$. Indeed, substituting some
term in a closed normal term yields the same term.\qed}
}
\VM{\proof{See the long version of the paper \cite{arrighidowek3}.}}
\VL{\begin{prop}
\label{substitution2}
If ${\bf v}_2 \lra_R^{*} {\bf w}_2$ then 
${\bf v}_1[{\bf v}_2/{\bf x}]
 \lra_R^{*} {\bf v}_1[{\bf w}_2/{\bf x}]$. 
\end{prop}
\proof{The reduction is a congruence.\qed}}}

\VML{
\begin{prop}
\label{commut1}
If ${\bf t} = {\bf u}$ or ${\bf t} \lra_R {\bf u}$ and if
${\bf t}\lra_{B}^{\parallel} {\bf v}$
then there exists ${\bf w}$ such that
${\bf u}\lra_{B}^{\parallel} {\bf w}$ and 
${\bf v}\lra_R^{*} {\bf w}$.
\end{prop}
\VML{\proof{
By induction on the structure of ${\bf t}$. If ${\bf t} = {\bf u}$ we
just take ${\bf w} = {\bf v}$.  Thus we focus in the rest of the proof
to the case where ${\bf t} \lra_R {\bf u}$. 

If the $B^{\parallel}$-reduction takes place at the
toplevel, then ${\bf t} = (\lambda {\bf x}\, {\bf t}_1)~{\bf t}_2$,
${\bf t}_2$ is a base vector and there exists terms ${\bf v}_1$ and
${\bf v}_2$ such that ${\bf t}_1 \lra_{B}^{\parallel} {\bf v}_1$,
${\bf t}_2 \lra_{B}^{\parallel} {\bf v}_2$ and ${\bf v} = {\bf
v}_1[{\bf v}_2/{\bf x}]$.  Neither $\lambda {\bf x}\, {\bf t}_1$ nor
${\bf t}_2$ is a sum, a product by a scalar or the null vector, hence
the $R$-reduction is just an application of the congruence thus there
exists terms ${\bf u}_1$ and ${\bf u}_2$ such that and ${\bf t}_1
\lra_R^{?} {\bf u}_1$ and ${\bf t}_2 \lra_R^{?} {\bf u}_2$.  Since
${\bf t}_2$ is a base vector, ${\bf u}_2$ is also a base vector.  By
induction hypothesis, there exist terms ${\bf w}_1$ and ${\bf w}_2$
such that ${\bf u}_1 \lra_{B}^{\parallel} {\bf w}_1$, ${\bf v}_1
\lra_R^{*} {\bf w}_1$, ${\bf u}_2 \lra_{B}^\parallel {\bf w}_2$ and
${\bf v}_2 \lra_R^{*} {\bf w}_2$.  We take ${\bf w} = {\bf w}_1[{\bf
w}_2/{\bf x}]$. We have $(\lambda {\bf x}\, {\bf u}_1)~{\bf u}_2
\lra_{B}^{\parallel} {\bf w}$ and by Proposition \ref{substitution}
\VL{and \ref{substitution2}} we also have ${\bf v}_1[{\bf v}_2/{\bf x}]
\lra_R^{*} {\bf w}$.

If the $R$-reduction takes place at the toplevel, we have to
distinguish several cases according to the rule used for this reduction.
\VL{\begin{itemize}
\item 
If ${\bf t} = {\bf t}_1 + {\bf 0}$ and ${\bf u} = {\bf t}_1$, then
there exists a term ${\bf v}_1$ such that ${\bf t}_1 
\lra_{B}^{\parallel} {\bf v}_1$ and 
${\bf v} = {\bf v}_1 + {\bf 0}$. We take ${\bf w} = {\bf v}_1$.

\item If ${\bf t} = 0 . {\bf t}_1$ and ${\bf u} = {\bf 0}$, then there
exists a term ${\bf v}_1$ such that  
${\bf t}_1 \lra_{B}^{\parallel} {\bf v}_1$ and 
${\bf v} = 0 . {\bf v}_1$. We take ${\bf w} = {\bf 0}$. 

\item If ${\bf t} = 1 . {\bf t}_1$ and ${\bf u} = {\bf t}_1$, then
there exists a term ${\bf v}_1$ such that ${\bf t}_1
\lra_{B}^{\parallel} 
{\bf v}_1$ and ${\bf v} = 1 . {\bf v}_1$. 
We take ${\bf w} = {\bf v}_1$. 

\item If ${\bf t} = \alpha . {\bf 0}$ and ${\bf u} = {\bf 0}$, 
then ${\bf v} = {\bf t}$. We take ${\bf w} = {\bf 0}$. 

\item If ${\bf t} = \alpha . (\beta . {\bf t}_1)$ 
and ${\bf u} = (\alpha \times \beta) . {\bf t}_1$, then 
there exists a term ${\bf v}_1$ such that ${\bf t}_1 \lra_{B}^{\parallel}
{\bf v}_1$ and ${\bf v} = \alpha . (\beta . {\bf v}_1)$.
We take ${\bf w} = (\alpha \times \beta) . {\bf v}_1$. 

\item If ${\bf t} = \alpha . ({\bf t}_1 + {\bf t}_2)$ and 
${\bf u} = \alpha . {\bf t}_1 + \alpha . {\bf t}_2 $, then 
there exist terms ${\bf v}_1$ and ${\bf v}_2$ such that 
${\bf t}_1  \lra_{B}^{\parallel} {\bf v}_1$, 
${\bf t}_2  \lra_{B}^{\parallel} {\bf v}_2$ and
${\bf v} = \alpha . ({\bf v}_1 + {\bf v}_2)$.
We take ${\bf w} = \alpha . {\bf v}_1 + \alpha . {\bf v}_2$. 

\item 
If ${\bf t} = \alpha . {\bf t}_1 + \beta . {\bf t}_1$ 
and ${\bf u} =  (\alpha + \beta). {\bf t}_1$, then 
by $(*)$ ${\bf t}_1$ is $L$-normal, thus ${\bf v} = {\bf t}$. 
We take ${\bf w} = {\bf u}$. 

The cases of the two other factorisation rules are similar.

\item 
If ${\bf t} = ({\bf t}_1 + {\bf t}_2)~{\bf t}_3$ and 
${\bf u} = {\bf t}_1 ~{\bf t}_3 + {\bf t}_2 ~{\bf t}_3$, then 
by $(**)$ the term ${\bf t}_1 + {\bf t}_2$ is $L$-normal. There exists
a term ${\bf v}_3$ such that ${\bf t}_3 \lra_{B}^{\parallel} {\bf
v}_3$ and ${\bf v} = ({\bf t}_1 + {\bf t}_2) ~{\bf v}_3$. 
We take ${\bf w} = {\bf t}_1 ~{\bf v}_3 + {\bf t}_2 ~{\bf v}_3$.

\item 
If ${\bf t} = (\alpha . {\bf t}_1) ~{\bf t}_2$ and
${\bf u} = \alpha.({\bf t}_1 ~{\bf t}_2)$, then by $(***)$ ${\bf t}_1$
is $L$-normal. There exists 
a term ${\bf v}_2$ such that ${\bf t}_2 \lra_{B}^{\parallel} {\bf
v}_2$ and ${\bf v} = (\alpha . {\bf t}_1) ~{\bf v}_2$. 
We take ${\bf w} = \alpha . ({\bf t}_1 ~{\bf v}_2)$.

\item 
If ${\bf t} = {\bf 0} ~{\bf t}_2$ and ${\bf u} = {\bf 0}$, then 
there exists a term ${\bf v}_2$ such that ${\bf t}_2
\lra_{B}^{\parallel} {\bf v}_2$ and 
${\bf v} = {\bf 0} ~{\bf v}_2$. We take ${\bf w} = {\bf 0}$.

\end{itemize}}\VM{See the details in the long version of the paper \cite{arrighidowek3}.}

\noindent The three other cases where a rule of group $A$ is applied are
symmetric.

Finally if both reductions are just applications
of the congruence we apply the induction hypothesis to the
subterms.\qed}}}\medskip

\begin{prop}[$\lra_R^{*}$ 
commutes with $\lra_{B}^{\parallel}$]~\\
\label{commut}
If ${\bf t} \lra_R^{*} {\bf u}$ and  
${\bf t}\lra_{B}^{\parallel} {\bf v}$
then there exists ${\bf w}$ such that
${\bf u}\lra_{B}^{\parallel} {\bf w}$ and 
${\bf v}\lra_R^{*} {\bf w}$.
\end{prop}
\VML{\proof{By induction on the length of the $\lra_R^{*}$ derivation.
If ${\bf t} = {\bf u}$ then we take ${\bf w} = {\bf v}$. Otherwise
there exists a term ${\bf t}_1$ such that ${\bf t} \lra_R {\bf t}_1
\lra_R^{*} {\bf u}$ with a shorter reduction from ${\bf t}_1$ to ${\bf
u}$.  Using Proposition \ref{commut1}, there exists a term ${\bf w}_1$
such that ${\bf t}_1 \lra_B^{\parallel} {\bf w}_1$ and ${\bf v}
\lra_R^{*} {\bf w}_1$. By induction hypothesis, there exists a term
${\bf w}$ such that ${\bf u} \lra_B^{\parallel} {\bf w}$ and ${\bf
w}_1 \lra_R^{*} {\bf w}$.  We have ${\bf u} \lra_B^{\parallel} {\bf
w}$ and ${\bf v} \lra_R^{*} {\bf w}$.\qed}}

\VML{
\begin{prop}[Substitution for $B^\parallel$]\label{substitution3}~\\
If ${\bf t}
\lra_{B}^{\parallel} 
{\bf t'}$ and ${\bf b} \lra_{B}^{\parallel} {\bf b'}$ then ${\bf t}[{\bf b}/{\bf x}] 
\lra_{B}^{\parallel} {\bf b'}[{\bf b'} /{\bf x}]$. Here ${\bf b}$ denotes a base vector.
\end{prop}
\VL{\proof{
By induction on the structure of ${\bf t}$.
\begin{itemize}
\item If ${\bf t}={\bf x}$ then ${\bf t'}={\bf x}$ and hence 
${\bf t}[{\bf b}/{\bf x}]={\bf b} 
\lra_{B}^{\parallel} {\bf b'}={\bf t'}[{\bf b'}/{\bf x}]$.
\item If ${\bf t}={\bf y}$ then ${\bf t'}={\bf y}$ and hence 
${\bf t}[{\bf b}/{\bf x}]={\bf y}={\bf t'}[{\bf b'}/{\bf x}]$.
\item If ${\bf t}=\lambda {\bf y}\,{\bf t}_1$ the $B^\parallel$-reduction
is just an application of the congruence. 
We have ${\bf t'}=\lambda {\bf y}\,.{\bf t_1'}$ with
${\bf t}_1\lra_{B}^{\parallel}{\bf t_1'}$ and the induction hypothesis
tells us that ${\bf t}_1[{\bf b}/{\bf x}]\lra_{B}^{\parallel}
{\bf t_1'}[{\bf b'}/{\bf x}]$. 
Hence  ${\bf t}[{\bf b}/{\bf x}]=\lambda {\bf y}\,{\bf t}_1[{\bf b}/{\bf x}] 
\lra_{B}^{\parallel} \lambda {\bf y}\,{\bf t_1'}[{\bf b'}/{\bf x}]={\bf t'}[{\bf b'}/{\bf x}]$.

\item If ${\bf t}=({\bf t}_1 ~ {\bf t}_2)$ then we consider two cases.
\begin{itemize}
\item We have ${\bf t}_1=\lambda{\bf y}~{\bf t}_3$, ${\bf t}_2$ a base state, and ${\bf t'}={\bf t_3'}[{\bf t_2'}/{\bf x}]$, i.e., a $B$-reduction occurs at top-level. 
By induction hypothesis we know that 
${\bf t}_3[{\bf b} /{\bf x}]\lra_{B}^{\parallel} {\bf t_3'}[{\bf b'} /{\bf x}]$ and 
${\bf t}_2[{\bf b} /{\bf x}]\lra_{B}^{\parallel} {\bf t_2'}[{\bf b'} /{\bf x}]$.
Because ${\bf t_2}$ and ${\bf b}$ are base vectors, so is ${\bf t}_2[{\bf b} /{\bf x}]$.
Hence ${\bf t}[{\bf b} /{\bf x}]=
(\lambda{\bf y}\,{\bf t}_3[{\bf b} /{\bf x}]){\bf t}_2[{\bf b} /{\bf x}]
\lra_{B}^{\parallel} {\bf t_3'}[{\bf b'} /{\bf x}][{\bf t_2'}[{\bf b'} /{\bf x}]/{\bf y}]=
{\bf t_3'}[{\bf t_2'}/{\bf y}][{\bf b'} /{\bf x}]={\bf t'}[{\bf b'} /{\bf x}]$.
\item If ${\bf t'}=({\bf t_1'} ~ {\bf t_2'})$ with ${\bf t}_1\lra_{B}^{\parallel}{\bf t_1'}$, 
${\bf t}_2\lra_{B}^{\parallel}{\bf t_2'}$, then by induction hypothesis we know that
${\bf t}_1[{\bf b} /{\bf x}]\lra_{B}^{\parallel} {\bf t_1'}[{\bf b'} /{\bf x}]$ and 
${\bf t}_2[{\bf b} /{\bf x}]\lra_{B}^{\parallel} {\bf t_2'}[{\bf b'} /{\bf x}]$.
Hence ${\bf t}[{\bf b} /{\bf x}]=({\bf t}_1[{\bf b} /{\bf x}]~{\bf t}_2[{\bf b} /{\bf x}])
\lra_{B}^{\parallel} ({\bf t_1'}[{\bf b'} /{\bf x}]~{\bf t_2'}[{\bf b'} /{\bf x}])={\bf t'}[{\bf b'} /{\bf x}]$.
\end{itemize}

\item If ${\bf t}={\bf 0}$ then ${\bf t'}={\bf 0}$ and hence 
${\bf t}[{\bf b} /{\bf x}]={\bf 0}={\bf t'}[{\bf b'} /{\bf x}]$.

\item If ${\bf t}$ is a sum the $B^\parallel$-reduction is just an application of the congruence. Therefore
${\bf t}$ is AC-equivalent to ${\bf t}_1+{\bf t}_2$ and ${\bf t'}$ is AC-equivalent to ${\bf t_1'}+{\bf t_2'}$ with ${\bf t}_1\lra_{B}^{\parallel}{\bf t_1'}$, 
${\bf t}_2\lra_{B}^{\parallel}{\bf t_2'}$. Then by induction hypothesis we know that
${\bf t}_1[{\bf b} /{\bf x}]\lra_{B}^{\parallel} {\bf t_1'}[{\bf b'} /{\bf x}]$ and 
${\bf t}_2[{\bf b} /{\bf x}]\lra_{B}^{\parallel} {\bf t_2'}[{\bf b'} /{\bf x}]$.
Hence ${\bf t}[{\bf b} /{\bf x}]={\bf t}_1[{\bf b} /{\bf x}]+{\bf t}_2[{\bf b} /{\bf x}]
\lra_{B}^{\parallel} {\bf t_1'}[{\bf b'} /{\bf x}]+{\bf t_2'}[{\bf b'} /{\bf x}]={\bf t'}[{\bf b'} /{\bf x}]$.

\item If ${\bf t}=\alpha.{\bf t}_1$ the $B^\parallel$-reduction is just an application of the congruence.
We have ${\bf t'}=\alpha.{\bf t_1'}$ with ${\bf t}_1\lra_{B}^{\parallel}{\bf t_1'}$ and the induction hypothesis tells us that ${\bf t}_1[{\bf b} /{\bf x}]\lra_{B}^{\parallel} {\bf t_1'}[{\bf b'} /{\bf x}]$.
Hence  ${\bf t}[{\bf b} /{\bf x}]=\alpha.{\bf t}_1[{\bf b} /{\bf x}]\lra_{B}^{\parallel} \alpha.{\bf t_1'}[{\bf b'} /{\bf x}]={\bf t'}[{\bf b'} /{\bf x}]$.\qed
\end{itemize}}}
\VM{\proof{
The proof is by induction on the structure of ${\bf t}$, and differs
very little from that of the classical version of this result. The
detail can be found in
the long version of the paper \cite{arrighidowek3}.}}}

\begin{prop}[Strong confluence of $B^\parallel$]\label{Bconfluence}~\\ 
If ${\bf t}\lra_{B}^{\parallel} {\bf u}$ and ${\bf t} \lra_{B}^{\parallel} {\bf v}$ 
then there exists ${\bf w}$ such that ${\bf u}\lra_{B}^{\parallel} {\bf w}$ and ${\bf v} \lra_{B}^{\parallel} {\bf w}$.
\end{prop}
\VL{\proof{
By induction on the structure of ${\bf t}$.

\begin{itemize}
\item If ${\bf t}$ is a variable then ${\bf u} =
  {\bf t}$ and ${\bf v} = {\bf t}$. We take ${\bf w} =
  {\bf t}$. 
\item If ${\bf t} = {\bf 0}$ then ${\bf u} =
  {\bf 0}$ and ${\bf v} = {\bf 0}$. We take ${\bf w} =
  {\bf 0}$. 
\item If ${\bf t} = \lambda {\bf x}~{\bf t}_1$ then 
 ${\bf u} = \lambda {\bf x}~{\bf u}_1$ 
with ${\bf t}_1 \lra_{B}^{\parallel} {\bf u}_1$
and ${\bf v} = \lambda {\bf x}~{\bf v}_1$ 
with ${\bf t}_1 \lra_{B}^{\parallel} {\bf v}_1$. 
By induction hypothesis, there exists a ${\bf w}_1$ such that 
${\bf u}_1\lra_{B}^{\parallel} {\bf w}_1$ and 
${\bf v}_1\lra_{B}^{\parallel} {\bf w}_1$. We take 
${\bf w} = \lambda {\bf x}~{\bf w}_1$. 
\item  If ${\bf t} = ({\bf t}_1~{\bf t}_2)$ then we consider 
two cases.
\begin{itemize}
\item If the term ${\bf t}_1$ has the form $\lambda {\bf x}
~ {\bf t}_3$ and ${\bf t}_2$ is a base vector.
We consider three subcases, according to the form of the
$B^{\parallel}$-reductions. 
Either 
${\bf v} = ({\bf v}_1~{\bf v}_2)$ 
with 
${\bf t}_1 \lra_{B}^{\parallel} {\bf v}_1$, 
${\bf t}_2 \lra_{B}^{\parallel} {\bf v}_2$, 
and 
${\bf u} = ({\bf u}_1~{\bf u}_2)$
with 
${\bf t}_1 \lra_{B}^{\parallel} {\bf u}_1$, 
${\bf t}_2 \lra_{B}^{\parallel} {\bf u}_2$. 
By induction hypothesis, there exists terms 
${\bf w}_1$ and ${\bf w}_2$ such that 
${\bf u}_1 \lra_{B}^{\parallel} {\bf w}_1$, 
${\bf v}_1 \lra_{B}^{\parallel} {\bf w}_1$, 
${\bf u}_2 \lra_{B}^{\parallel} {\bf w}_2$,
${\bf v}_2 \lra_{B}^{\parallel} {\bf w}_2$.
We take ${\bf w} = ({\bf w}_1~{\bf w}_2)$.

Or ${\bf v} = {\bf v}_3[{\bf v}_2/{\bf x}]$ with
${\bf t}_3\lra_{B}^{\parallel} {\bf v}_3$, ${\bf t}_2 \lra_{B}^{\parallel}
{\bf v}_2$, and ${\bf u}=((\lambda{\bf x}~{\bf u}_3)~{\bf u}_2)$ with
${\bf t}_3 \lra_{B}^{\parallel} {\bf u}_3$, ${\bf t}_2 \lra_{B}^{\parallel}
{\bf u}_2$.  Since ${\bf t}_2$ is a base vector, ${\bf u}_2$ and ${\bf v}_2$ are also
base vectors.  By induction hypothesis, there exist terms ${\bf w}_3$
and ${\bf w}_2$ such that ${\bf u}_3 \lra_{B}^{\parallel} {\bf w}_3$,
${\bf v}_3 \lra_{B}^{\parallel} {\bf w}_3$, ${\bf u}_2 \lra_{B}^{\parallel}
{\bf w}_2$, ${\bf v}_2 \lra_{B}^{\parallel} {\bf w}_2$.  We take ${\bf w} =
{\bf w}_3[{\bf w}_2/{\bf x}]$. We have $(\lambda
{\bf x}\,{\bf u}_3)~{\bf u}_2 \lra_{B}^{\parallel}
{\bf w}_3[{\bf w}_2/{\bf x}]$ by definition of $B^\parallel$.  By
Proposition \ref{substitution3} we also have
${\bf v}_3[{\bf v}_2/{\bf x}] \lra_{B}^\parallel {\bf w}_3[{\bf w}_2/{\bf x}]$.

Or ${\bf v} = {\bf v}_3[{\bf v}_2/{\bf x}]$ 
with 
${\bf t}_3\lra_{B}^{\parallel} {\bf v}_3$, 
${\bf t}_2 \lra_{B}^{\parallel} {\bf v}_2$, 
and 
${\bf u}== {\bf u}_3[{\bf u}_2/{\bf x}]$
with
${\bf t}_3 \lra_{B}^{\parallel} {\bf u}_3$,
${\bf t}_2 \lra_{B}^{\parallel} {\bf u}_2$.  
Since ${\bf t}_2$ is a base vector, 
${\bf u}_2$ and ${\bf v}_2$  are base vectors also.
By induction hypothesis, there exist terms 
${\bf w}_3$ and ${\bf w}_2$ such that 
${\bf u}_3 \lra_{B}^{\parallel} {\bf w}_3$, 
${\bf v}_3 \lra_{B}^{\parallel} {\bf w}_3$, 
${\bf u}_2 \lra_{B}^{\parallel} {\bf w}_2$,
${\bf v}_2 \lra_{B}^{\parallel} {\bf w}_2$.
We take ${\bf w} = {\bf w}_3[{\bf w}_2/{\bf x}]$.  
By Proposition \ref{substitution3} 
we have both ${\bf u}_3[{\bf u}_2/{\bf x}] \lra_{B}^\parallel {\bf w}_3[{\bf w}_2/{\bf x}]$
and ${\bf v}_3[{\bf v}_2/{\bf x}] \lra_{B}^\parallel {\bf w}_3[{\bf w}_2/{\bf x}]$.

\item Otherwise the $B^{\parallel}$-reduction is just an application
of the congruence, i.e., ${\bf v} = ({\bf v}_1~{\bf v}_2)$ 
with 
${\bf t}_1 \lra_{B}^{\parallel} {\bf v}_1$, 
${\bf t}_2 \lra_{B}^{\parallel} {\bf v}_2$, 
and 
${\bf u} = ({\bf u}_1~{\bf u}_2)$
with 
${\bf t}_1 \lra_{B}^{\parallel} {\bf u}_1$, 
${\bf t}_2 \lra_{B}^{\parallel} {\bf u}_2$. 
By induction hypothesis, there exists terms 
${\bf w}_1$ and ${\bf w}_2$ such that 
${\bf u}_1 \lra_{B}^{\parallel} {\bf w}_1$, 
${\bf v}_1 \lra_{B}^{\parallel} {\bf w}_1$, 
${\bf u}_2 \lra_{B}^{\parallel} {\bf w}_2$,
${\bf v}_2 \lra_{B}^{\parallel} {\bf w}_2$.
We take ${\bf w} = ({\bf w}_1~{\bf w}_2)$.

\end{itemize}

\item If ${\bf t}$ is a sum then 
the $B^{\parallel}$-reduction is just an application
of the congruence. The term ${\bf t}$ is AC-equivalent to a sum 
${\bf t}_1 + {\bf t}_2$, 
the term ${\bf u}$ is AC-equivalent to a sum
${\bf u}_1 + {\bf u}_2$ with
${\bf t}_1 \lra_{B}^{\parallel} {\bf u}_1$, 
${\bf t}_2 \lra_{B}^{\parallel} {\bf u}_2$,
and the term ${\bf v}$ is AC-equivalent to a sum 
${\bf v}_1 + {\bf v}_2$ such that 
${\bf t}_1 \lra_{B}^{\parallel} {\bf v}_1$ and 
${\bf t}_2 \lra_{B}^{\parallel} {\bf v}_2$.
By induction hypothesis, there exist terms 
${\bf w}_1$ and ${\bf w}_2$ such that 
${\bf u}_1 \lra_{B}^{\parallel} {\bf w}_1$, 
${\bf v}_1 \lra_{B}^{\parallel} {\bf w}_1$, 
${\bf u}_2 \lra_{B}^{\parallel} {\bf w}_2$,
${\bf v}_2 \lra_{B}^{\parallel} {\bf w}_2$.
We take ${\bf w} = {\bf w}_1 + {\bf w}_2$. 

\item If finally, ${\bf t} = \alpha . {\bf t}_1$ then
the $B^{\parallel}$-reduction is just an application
of the congruence. 
We have ${\bf u} = \alpha . {\bf u}_1$ with 
${\bf t}_1 \lra_{B}^{\parallel} {\bf u}_1$, 
and ${\bf v} = \alpha {\bf v}_1$ with 
${\bf t}_1 \lra_{B}^{\parallel} {\bf v}_1$. 
By induction hypothesis, there exists a term ${\bf w}_1$ such that 
${\bf u}_1\lra_{B}^{\parallel} {\bf w}_1$, 
${\bf v}_1\lra_{B}^{\parallel} {\bf w}_1$. 
We take  ${\bf w} = \alpha . {\bf w}_1$. \qed
\end{itemize}}}
\VM{\proof{The proof is by induction on the structure of ${\bf t}$, and
  differs very little from that of the classical version of this
  result. The detail can be found in the long version of the paper \cite{arrighidowek3}.}}

\begin{prop}[Hindley-Rosen lemma]
\label{HR}
If the relations $X$ and $Y$ are strongly confluent and commute then
the relation $X \cup Y$ is confluent. 
\end{prop}

\begin{thm}\label{thconfluence}
The system $L$ is confluent.
\end{thm}

\proof{
By Proposition \ref{Rconf}, the relation $\lra_R$ is confluent,
hence $\lra_R^*$ is strongly confluent.
By Proposition \ref{Bconfluence}, the relation $\lra_B^{\parallel}$ 
is strongly confluent. 
By Proposition \ref{commut}, the relations 
$\lra_R^*$ and $\lra_B^{\parallel}$ commute. Hence, by Proposition 
\ref{HR} the relation $\lra_R^* \cup \lra_B^{\parallel}$ is confluent.
Hence, the relation $\lra_L$ is confluent.\qed}

\VML{
\begin{cor}[No-cloning in the Linear-algebraic $\lambda$-calculus]
There is no term \textsc{Clone} such that for all term ${\bf v}$,
$(\textsc{Clone}~{\bf v}) \lra_L^* ({\bf v}\otimes {\bf v})$.
\end{cor}
\proof{
Note that $\otimes$, ${\bf true}$ and ${\bf false}$ stand for
the terms introduced in Section \ref{encodings}.
Say $(\textsc{Clone}~{\bf v}) \lra_L^* ({\bf v}\otimes {\bf v})$ for all ${\bf v}$.
Let ${\bf v}=\alpha.{\bf true}+\beta.{\bf false}$ be in closed normal form.
Then by the $A$-rules we have 
$(\textsc{Clone}~(\alpha.{\bf true}+\beta.{\bf false})) \lra_L^* \alpha.(\textsc{Clone}~{\bf true})+\beta.(\textsc{Clone}~{\bf false})$. Next, according to our supposition on \textsc{Clone}, this further reduces to $\alpha.({\bf true}\otimes{\bf true})$+$\beta.({\bf false}\otimes{\bf false})$. 
But our supposition on \textsc{Clone}, also says that $(\textsc{Clone}~(\alpha.{\bf true}+\beta.{\bf false}))$ reduces to $(\alpha.{\bf true}+\beta.{\bf false})\otimes(\alpha.{\bf true}+\beta.{\bf false})$. Moreover the two cannot be reconciled into a common reduct, because they are normal. 
Hence our supposition would break the confluence; it cannot hold.
}}
Note that $\lambda {\bf x}\,{\bf v}$ on the other hand can be
duplicated, because it is thought as the (plans of) the classical
machine for building ${\bf v}$ -- in other words it stands for
potential parallelism rather than actual parallelism. As expected
there is no way to transform ${\bf v}$ into $\lambda {\bf x}\, {\bf v}$ in general; confluence ensures that the calculus handles this distinction in a consistent manner.\qed

\section{Current works}\label{currentworks}

\subsection{Algebraic $\lambda$-calculus}

As we have mentioned in the introduction the idea of endowing the
$\lambda$-calculus with a vector space has emerged simultaneously and
independently in a different context. Indeed, the exponential-free
fragment of Linear Logic is a logic of resources where the
propositions themselves stand for those resources -- and hence cannot
be discarded nor copied. When seeking to find models of this logic,
one obtains a particular family of vector spaces and differentiable
functions over these. It is by trying to capture back these
mathematical structures into a programming language that T. Ehrhard
and L. Regnier have defined the {\em differential $\lambda$-calculus}
\cite{Ehrhard}, which has an intriguing differential operator as a
built-in primitive, and some notion of module of the
$\lambda$-calculus terms, over the natural numbers. More recently
L. Vaux \cite{Vaux} has focused his attention on a ``differential
$\lambda$-calculus without differential operator'', extending the
module to finitely splitting positive real numbers. He obtained a
confluence result in this case, which stands even in the untyped
setting. More recent works on this {\em Algebraic $\lambda$-calculus}
tend to consider arbitrary scalars \cite{EhrhardSystemF, Tasson}.
This Algebraic $\lambda$-calculus and the Linear-algebraic
$\lambda$-calculus we presented in this paper are very similar not
only in names: they both merge higher-order computation, be it
terminating or not, in its simplest and most general form (namely the
untyped $\lambda$-calculus) together with linear algebra in its
simplest and most general form also (the axioms of vector
spaces). Skipping over details a closer inspection unravels that:
\begin{itemize}
\item
the application in the Algebraic $\lambda$-calculus is left linear but
not right linear;
\item 
the abstraction in the Algebraic $\lambda$-calculus is a linear unary 
operator;
\item 
the rewriting is modulo vector space axioms, and these axioms are not
transformed into rewrite rules of the system.
\end{itemize}

\noindent It could be said the last two points are only minor
differences; design choices in some sense. Arguably those of {\em
  Lineal} are advantageous because they yield a more robust confluence
proof, valid for arbitrary scalars. If we lift these two differences,
{\em Lineal} simulates the Algebraic $\lambda$-calculus
\cite{ArrighiVauxmanuscript}. The first point is a more important
difference, with justification right within the origins of the
Algebraic $\lambda$-calculus and the Differential
$\lambda$-calculus. Recently, however, it has been shown that the
difference really amounts to a choice between call-by-name and
call-by-value oriented strategies. The encoding of one strategy into
another still works \cite{CurrentWorkonLinealintoAlgLam} --- hence it
could be said that the two calculi are essentially equivalent.

\subsection{Types}

Whilst terms in our calculus seem to form a vector space, the very
definition of a norm is difficult in our context: deciding whether a
term terminates is undecidable; but these terms produce infinities,
hence convergence of a vector space norm is undecidable.  Related to
this precise topic, L. Vaux has studied simply typed algebraic
$\lambda$-calculus, ensuring convergence of a vector space norm
\cite{Vaux}. Following his work, C. Tasson has studied some
model-theoretic properties of the {\em barycentric} ($\sum
\alpha_i=1$) subset of this simply typed calculus \cite{Tasson}. A
recent work by T. Ehrhard proves the convergence of a Taylor series
expansion of Algebraic $\lambda$-calculus terms, via a System $F$
typing system \cite{EhrhardSystemF}.\\ 
Hence, standard type systems ensure the convergence of the vector
space norm of a term. And indeed it is not so hard to define a simple
extension of System $F$ that fits {\em Lineal} --- just by providing
the needed rules to type additions, scalar products and the null
vector in some trivial manner, as we did in \cite{Scalar,Scalar2}. As
expected one obtains strong normalisation from this type system. An
important byproduct of this result is that one can then remove the
conditions $(*)-(***)$ that limit the reduction rules of {\em Lineal}
(see Section \ref{language}), because their purpose was really to keep
indefinite forms from reducing (such as ${\bf t}-{\bf t}$, with ${\bf
  t}$ not normal and hence potentially infinite). In other words types
make \emph{Lineal} into a simpler language.\\
Yet standard type systems are unable for instance to impose upon the
language that any well-typed linear combination of terms
$\sum\alpha_i.{\bf t}_i$ has $\sum \alpha_i=1$. That is unless they
are provided with a handle upon these scalars. This is the purpose of
the {\em scalar} type system which was recently proposed
\cite{Scalar,Scalar2}. This type system which manages to keep track of
``the amount of a type'' by summing the amplitudes of its contributing
terms, and reflects this amount within the type. As an example of its
uses, it was demonstrated that this provides a type system which
guarantees well-definiteness of probabilistic functions in the sense
that it specializes {\em Lineal} into a probabilistic, higher-order
$\lambda$-calculus. We are still looking for a type system that would
impose that linear combination of terms $\sum\alpha_i.{\bf t}_i$ have
$\sum |\alpha_i|^2=1$, as suited for quantum computing.

\subsection{Models}

The functions expressed in our language are linear operators upon the
space constituted by its terms. It is strongly inspired from the more
preliminary \cite{arrighidowek2}, where terms clearly formed a vector
space.  However because the calculus higher-order, we get forms of
infinities coming into the game.  Thus, the underlying algebraic
structure is not as obvious as in \cite{arrighidowek2}. Moreover one
can notice already that since the non-trivial models of the untyped
$\lambda$-calculus are all uncountable, the models of
(Linear-)Algebraic $\lambda$-calculus are likely to be vector spaces
having an uncountable basis.  These are fascinating, open questions,
but whose difficulty explain why we have not provided a denotational
semantics for {\em Lineal} in this paper.  This issue of models of
(Linear-)Algebraic $\lambda$-calculus is a challenging, active topic
of current research. We know of the categorical model of simply typed
{\em Lineal} with fixpoints \cite{Valironmanuscript}, which establishes
a connection between the canon and uncanon construct of Section 
\ref{encodings} and monads {\em \`a la} Moggi \cite{Moggi}.
The finiteness 
space model of
simply typed Algebraic $\lambda$-calculus \cite{finiteness,Tasson}
does not easily carry through to {\em Lineal}, which is call-by-value
oriented.  Recently, a syntactic finiteness space model of System $F$
algebraic $\lambda$-calculus has been developed in
\cite{EhrhardSystemF}.

\section{Conclusion}\label{conclusion}

\subsection{Summary} 

When merging the untyped $\lambda$-calculus with
linear algebra one faces two different problems. First of all
simple-minded duplication of a vector is a non-linear operation
(cloning) unless it is restricted to base vectors and later
extended linearly (copying). Second of all we can express
computable but nonetheless infinite series of vectors, hence yielding
some infinities and the troublesome indefinite forms. Here again this
is fixed by restricting the evaluation of these indefinite forms, this
time to normal vectors. Both problems show up when looking at the
confluence of the Linear-algebraic $\lambda$-calculus ({\em Lineal}).\\ 
The architecture of the proof of confluence seems well-suited to any
non-trivial rewrite systems having both some linear algebra and some
infinities as its key ingredients. 
Moreover the proof of confluence entails a no-cloning result for Lineal, 
in accordance with the linearity of quantum physics.

\VML{\smallskip}\noindent 
\subsection{Perspectives} 

{\em Lineal} merges
higher-order computation with linear algebra in a minimalistic 
manner. Such a foundational approach is also taking place for instance in
\cite{Abramsky} via some categorical formulations of quantum theory
exhibiting nice composition laws and normal forms, without explicit
states, fixed point or the possibility to replicate gate
descriptions. As for \cite{Abramsky} although we have shown that
quantum computation can be encoded in our language, {\em Lineal} remains some way apart from a
model of quantum computation, because it allows evolutions which are
not unitary. Establishing formal connections with this categorical approach 
does not seem an easy matter but is part of our objectives.\\
These connections might arise through typing. 
Finding a type system which specializes {\em Lineal} into a strictly quantum programming language (enforcing the unitary constraint) is not only our next step on the list, it is actually 
the principal aim and motivation for this work: we wish to extend the Curry-Howard
isomorphism between proofs/propositions and programs/types to a linear-algebraic, 
quantum setting. Having merged higher-order computation with linear-algebra in a minimalistic manner,
which does not depend on any particular type systems, grants us a complete liberty to now explore
different forms of this isomorphism. For instance we may expect different type systems to have different 
fields of application, ranging from fine-grained entanglement-analysis for quantum computation \cite{SimonEntanglement, ProstEntanglement},
to opening connections with linear logic \cite{Additive} or even giving rise to some novel, quantitative logics \cite{Scalar}.\\

\section*{Acknowledgments}

The authors would like to thank Alejandro D\'iaz-Caro, Evelyne Contejean,
Philippe Jorrand, Jean-Pierre Jouannaud, Claude March\'e, Simon 
Perdrix, Beno\^it Valiron and Lionel Vaux for some enlightening discussions.


{\small \begin{thebibliography}{99.}
\bibitem{AbramskyLL}
S. Abramsky, 
\emph{Computational Interpretations of Linear Logic}, {Theoretical Computer Science}, $\mathbf{111}$, {3--57}, (1993). 
\bibitem{Abramsky} S. Abramsky, B. Coecke, \emph{A categorical semantics of quantum protocols}
LICS, IEEE Computer Society, 415-425, (2004).
\bibitem{Adleman} L. Adleman, J. DeMarrais, M. Huang,
\emph{Quantum Computability}, SIAM J. on Comp., $\mathbf{26}$, 5,
1524-1540, (1997).
\bibitem{Altenkirch} 
T. Altenkirch, J. Grattage, J.K. Vizzotto, A. Sabry, An Algebra
of Pure Quantum Programming, {\em Third International Workshop on Quantum
Programming Languages}, Electronic Notes of Theoretical Computer
Science, 170C, 23-47, (2007).
\bibitem{arrighidowek1} P. Arrighi, G. Dowek, {\em A computational
definition of the notion of vector space},
ENTCS ${\bf 117}$, 249-261, (2005).
\bibitem{arrighidowek2} P. Arrighi, G. Dowek, {\em Linear-algebraic lambda-calculus}, 
in P. Selinger (Ed.), International workshop on quantum programming languages, Turku Centre
for Computer Science General Publication, {\bf 33}, 21-38, (2004).
\bibitem{arrighidowek3} P. Arrighi, G. Dowek, {\em Linear-algebraic lambda-calculus: higher-order, encodings, confluence}, arXiv:quant-ph/0612199.
\bibitem{arrighidowek4} P. Arrighi, G. Dowek, 
{\small \tt www-roc.inria.fr/who/Gilles.Dowek/Prog/lineal.html}.
\bibitem{251} P. Arrighi, G. Dowek, 
{\em On the critical pairs of a rewrite system for vector spaces}, available
on the web page of the authors, see {\small \tt 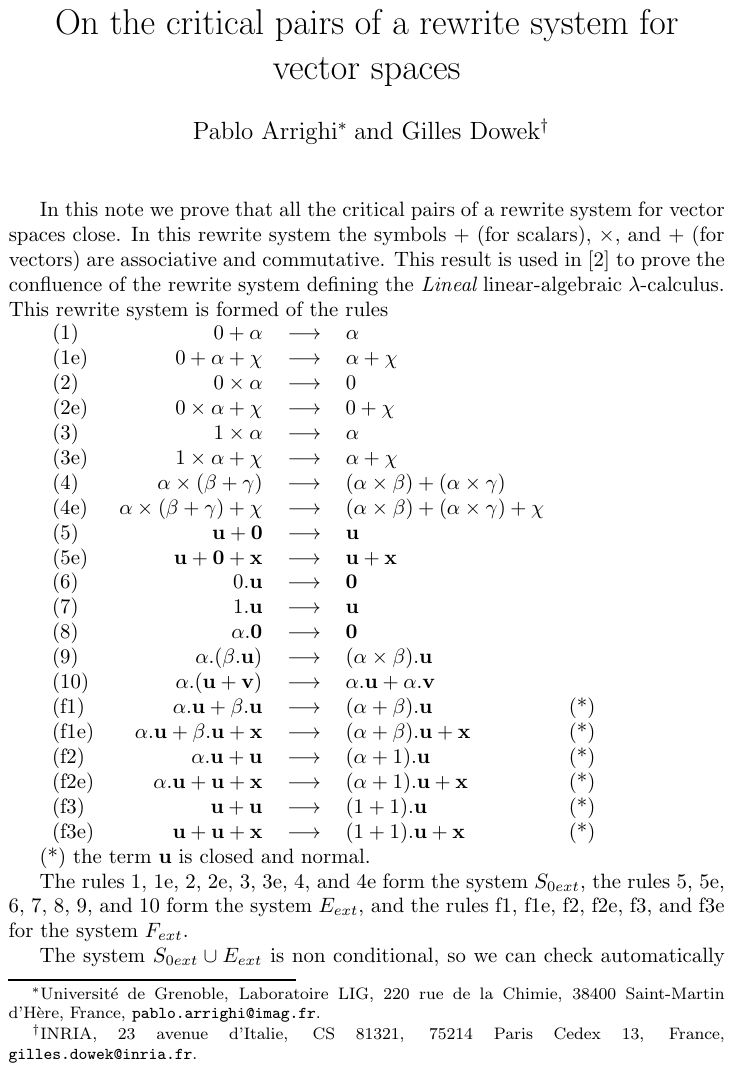}, (2012).
\bibitem{Scalar} P. Arrighi, A. D{\'i}az-Caro, \emph{Scalar System {F} for Linear-Algebraic $\lambda$-Calculus: Towards a Quantum Physical Logic}, {Proceedings of the 6th International Workshop on Quantum Physics and Logic}, ENTCS {206--215}, (2009).
\bibitem{Scalar2} P. Arrighi, A. D{\'i}az-Caro, \emph{A System {F} accounting for scalars}, Preprint: arXiv:0903.3741, (2009).
\bibitem{ArrighiVauxmanuscript}P. Arrighi, L. Vaux, \emph{Embedding Algebraic Lambda-calculus into Lineal}, Private communication, (2009).
\bibitem{Bernstein} E. Bernstein, U. Vazirani, \emph{Quantum Complexity Theory}, Annual ACM symposium on Theory of Computing, ${\bf 25}$, (1993).
\bibitem{Birkhoff} G. Birkhoff, {\em On the Structure of Abstract Algebras}, Proc. Cambridge Phil. Soc., ${\bf 31}$, (1935).
\bibitem{Boudol} G. Boudol, \emph{Lambda-calculi for (strict) parallel functions}, Information and Computation, ${\bf 108} (1)$, 51-127, (1994). 
\bibitem{Bournez} O. Bournez, M. Hoyrup, \emph{Rewriting Logic and Probabilities}, Rewriting Techniques and Applications, LNCS {\bf 2706}, (2003). 
\bibitem{Boykin} P. Boykin, T. Mor, M. Pulver, V. Roychowdhury, F.
Vatan, \emph{On universal and fault-taulerant quantum computing},
arxiv:quant-ph/9906054
\bibitem{ChiribellaValiron} G. Chiribella, G.  D'Ariano, P. Perinotti, B. Valiron, {\em Beyond Quantum Computers}, Arxiv preprint arXiv:0912.0195, (2009).
\bibitem{cime} 
{\em The CiME Rewrite Tool}, {\tt http://cime.lri.fr/}.
\bibitem{Cohen} D. Cohen, P. Watson, \emph{An efficient representation of arithmetic for term rewriting}, Proc. of the 4th Conference on Rewrite Techniques and Applications, LNCS
\bibitem{Dershowitz} N. Dershowitz, J.-P. Jouannaud, \emph{Rewrite
systems}, Handbook of theoretical computer science, Vol.
\textbf{B}: formal models and semantics, MIT press, (1991).
\bibitem{Deutsch} D. Deutsch, R. Josza, \emph{Rapid solution of problems by quantum
computation.} Proc. of the Roy. Soc. of London A, ${\bf 439}$,
553-558, (1992).
\bibitem{CurrentWorkonLinealintoAlgLam} A. D\'iaz-Caro, S. Perdrix, C. Tasson, B. Valiron \emph{Equivalence of Algebraic $\lambda$-calculi}, HOR 2010.
\bibitem{Additive} A. D\'iaz-Caro, B. Petit, {\em From Additive Logic to Linear Logic}, manuscript, (2010).
\bibitem{Dougherty} D. Dougherty, \emph{Adding Algebraic Rewriting to the Untyped Lambda Calculus}, Proc. of the Fourth International Conference on Rewriting Techniques and Applications, 1992.
\bibitem{Ehrhard} T. Ehrhard, L. Regnier, \emph{The differential lambda-calculus}, 
Theoretical Computer Science, ${\bf 309}$, 1--41, (2003).
\bibitem{finiteness} T. Ehrhard, \emph{Finiteness spaces}, Mathematical Structures in Computer Science, {\bf 15}(4), 615--646, (2005).
\bibitem{EhrhardSystemF} T. Ehrhard, \emph{A finiteness structure on resource terms}, LICS 2010, to appear.
\bibitem{FernandezMackie} M. Fernandez and I. Mackie, {\em Closed
Reductions in the $\lambda$-calculus}, Computer Science Logic,
Lecture Notes in Computer Science 1683, (1999).
\bibitem{Hankin} A. Di Pierro, C. Hankin, H. Wiklicky, \emph{Probabilistic $\lambda$-calculus and quantitative program analysis}, J. of Logic and Computation, ${\bf 15}(2)$, 159-179, (2005).
\bibitem{Gay} S. J. Gay, {\em Quantum programming languages: survey and bibliography}, Mathematical Structures in Computer Science, {\bf 16}(4), 581--600, (2006).
\bibitem{Girard1} J.-Y. Girard. \emph{Linear logic.} Theoretical Computer Science,
${\bf 50}$, 1-102, (1987).
\bibitem{Grover} {L. K. Grover}, \emph{Quantum Mechanics Helps in Searching for a Needle in a Haystack},
{Phys. Rev. Lett.}, ${\bf 79}(2)$, {325--328}, (1997).
\bibitem{Catuscia} O. M. Herescu, C. Palamidessi, \emph{Probabilistic asynchronous pi-calculus}, ETAPS, LNCS ${\bf 1784}$, 146--160, (2000).
\bibitem{KnuthBendix} G. Huet, {\em A complete proof of correctness of the Knuth-Bendix completion algorithm}, Journal of Computer and System Sciences, {\bf 23}(1), pages 11--21, (1981).
\bibitem{JouannaudKirchner} J.-P. Jouannaud,  H. Kirchner, 
\emph{Completion of a Set of Rules Modulo a Set of Equations}, 
SIAM J. of Computing, ${\bf 15}(4)$, 1155--1194, (1986).
\bibitem{Kitaev} A. Kitaev, \emph{Quantum computation, algorithms
and error correction}, Russ. Math. Surv., $\mathbf{52}$, 6,
1191-1249, (1997).
\bibitem{Moggi} E. Moggi, \emph{Notions of computation and monads}, Information
and Computation, {\bf 93}, 55--92, (1991).
\bibitem{Newman} M. H. A. Newman, \emph{On theories with a combinatorial definition of "equivalence"}, Annals of Mathematics, $\mathbf{43}2$, 223--243, (1942). 
\bibitem{Nielsen} M. A. Nielsen, \emph{Universal quantum computation using only projective measurement,
quantum memory, and preparation of the 0 state}, Phys. Rev. A, ${\bf 308}$, 96-100, (2003).
\bibitem{OreshkovCostaBrukner}, O. Oreshkov, F. Costa, C. Brukner, {\em Quantum correlations with no causal order}, {Arxiv preprint arXiv:1105.4464}, (2011).
\bibitem{PetersonStickel}  G. E. Peterson, M. E. Stickel, 
\emph{Complete Sets of Reductions for Some Equational Theories},
{J. ACM}, ${\bf 28}(2)$, {233-264}, (1981).
\bibitem{Perdrix} S. Perdrix, \emph{State transfer instead of teleportation in measurement-based quantum
computation}, Int. J. of Quantum Information , ${\bf 1}(1)$, 219-223, (2005).
\bibitem{SimonEntanglement}  S. Perdrix,
\emph{Quantum entanglement analysis based on abstract interpretation},
 SAS 2008, LNCS {\bf 5079}, (2008).
\bibitem{ProstEntanglement} F. Prost, C. Zerrari, \emph{Reasoning about Entanglement and Separability in Quantum Higher-Order Functions}, UC 2008, Proceedings of the 8th International Conference on Unconventional Computation, 219--235 (2009).
\bibitem{Briegel} {R. Raussendorf,  D.E. Browne, H.J. Briegel}, \emph{The one-way quantum computer - a non-network model of quantum computation}, {Journal of Modern Optics}, ${\bf 49}$, p. 1299, (2002).
\bibitem{Rudolph} T. Rudolph, L. Grover, \emph{A two rebit gate universal for quantum
computing}, october 2002, arxiv:quant-ph/0210187.
\bibitem{Selinger} P. Selinger, \emph{Towards a quantum programming language}, 
Math. Struc. in Computer Science, ${\bf 14}(4)$, 527-586, (2004).
\bibitem{Tasson} C. Tasson, \emph{Algebraic Totality, towards Completeness}, TLCA 2009: Proceedings of the 9th International Conference on Typed Lambda Calculi and Applications, 325--340, (2009).
\bibitem{Valiron} P. Selinger, B. Valiron, \emph{A lambda calculus for quantum computation with classical control}, Math. Struc. in Computer Science, ${\bf 16}(3)$, 527-552, (2006).
\bibitem{Valironmanuscript} B. Valiron, 
{\em A Typed, Algebraic, Computational Lambda-Calculus.}
Mathematical Structures in Computer Science (to appear).
\bibitem{Shor} P. W. Shor,  \emph{Polynomial-Time Algorithms for Prime Factorization and Discrete Logarithms on a Quantum Computer}, SIAM J. on Computing, ${\bf 26}$, 1484-1509, (1997).
\bibitem{Solovay1} R. Solovay, manuscript, (1995).
\bibitem{Solovay2} R. Solovay, A. Yao, \emph{Quantum Circuit Complexity and
Universal Quantum Turing Machines}, manuscript, (1996).
\bibitem{VanTonder1} A.~Van Tonder, \emph{A Lambda Calculus for Quantum Computation}, july 2003,
arXiv:quant-ph/0307150.
\VML{\bibitem{VanTonder2} A.~Van Tonder, \emph{Quantum Computation, Categorical Semantics and Linear
Logic}, december 2003, arXiv:quant-ph/0312174.}
\bibitem{Vaux} L. Vaux, \emph{On linear combinations of lambda-terms}, Proceedings of RTA 2007, LNCS ${\bf 4533}$, (2007). 
\bibitem{Walters} H. Walters, H. Zantema, \emph{Rewrite systems for integer
arithmetic}, Proc. of Rewriting Techniques and Applications 94,
6th Int. Conf., LNCS $\mathbf{914}$, 324-338, (1995).
\bibitem{cloning} W. K. Wooters, W. H. Zurek, \emph{A single quantum cannot be cloned}, Nature ${\bf 299}$, 802-803, (1982).
\end{thebibliography}}
\end{document}